\documentclass{article}

\usepackage{arxiv}

\usepackage[utf8]{inputenc} % allow utf-8 input
\usepackage[T1]{fontenc}    % use 8-bit T1 fonts
\usepackage{hyperref}       % hyperlinks
\usepackage{url}            % simple URL typesetting
\usepackage{booktabs}       % professional-quality tables
\usepackage{amsfonts}       % blackboard math symbols
\usepackage{nicefrac}       % compact symbols for 1/2, etc.
\usepackage{microtype}      % microtypography
\usepackage{graphicx}
\usepackage{doi}

\title{Nonparametric Estimation of a distribution function from doubly truncated data under dependence}

\author{\href{https://orcid.org/0000-0002-0570-0650}{\includegraphics[scale=0.06]{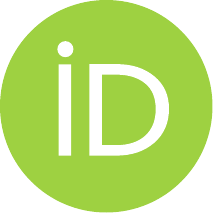}\hspace{1mm}Carla Moreira}\\%\thanks{Use footnote for providing further
	Centre of Mathematics\\
	University of Minho\\
   	Braga - Portugal \\
    SiDOR research group\\
    University of Vigo - Spain\\
	\texttt{carlamgmm@gmail.com} \\
	\And
	Jacobo de U\~na - \'Alvarez \\
	Department of Statistics and Operations Research\\
    SiDOR research group \& CINBIO
	University of Vigo\\
	Vigo- Spain \\
	\texttt{jacobo@uvigo.es} \\
	 \And
	 Roel Braekers \\
	 I-BioStat, Hasselt University \\
	 Agoralaan, B-3590 Diepenbeek, Belgium \\
	\texttt{roel.braekers@uhasselt.be } \\
}

\renewcommand{\shorttitle}{\textit{arXiv} Template}

\hypersetup{
pdftitle={A template for the arxiv style},
pdfsubject={q-bio.NC, q-bio.QM},
pdfauthor={David S.~Hippocampus, Elias D.~Striatum},
pdfkeywords={First keyword, Second keyword, More},
}

\begin{document}
\maketitle

\begin{abstract}
The NPMLE of a distribution function from doubly truncated data was introduced in the seminal
paper of Efron and Petrosian \cite{Efron99}. The consistency of the Efron-Petrosian estimator depends however on the
assumption of independent truncation. In this work we introduce an extension of the Efron-Petrosian NPMLE when
the variable of interest and the truncation variables may be dependent. The proposed estimator is constructed on the basis of a copula function which represents the dependence structure between the variable of interest and the truncation variables. Two different iterative algorithms to
compute the estimator in practice are introduced, and their performance is explored through an intensive Monte Carlo simulation study. We illustrate the use of the estimators on two real data examples.
\end{abstract}

% keywords can be removed

\section{Introduction}
\label{sec:Section1}

Censored and truncated data appear in fields like Astronomy, Epidemiology or Survival
analysis, among others. For instance, censored survival data appear
because of lost to follow-up cases or due to time limitations in the following-up
of the individuals. A common censoring pattern that appears in many medical applications is the so called interval censoring. Such type of censoring  arises when the occurrence of the final event of interest cannot be exactly observed and the
inter-event time is only known to lie in an interval \cite{Gentleman1994}, \cite{Lindsey98}.  Random
truncation occurs when only event times falling on a given time interval (which varies from individual to individual) can be observed. Left-truncation and right-truncation are special truncation
patterns emerging when the truncation interval is unbounded from above (left-truncation) or from
below (right-truncation). Both issues of censoring and truncation may introduce a systematic bias
in estimation, so specific methods to overcome these problems must be applied. This paper is focused on random double truncation.

Let $X^{\ast }$  be the random variable of ultimate interest, the target variable, with distribution function (df)
$F$, and assume that it is doubly truncated by the random pair
$\left( U^{\ast },V^{\ast }\right) $ with joint df $G$, where
$U^{\ast }$ and $V^{\ast }$ ($U^{\ast }\leq V^{\ast }$) are the left
and right truncation variables respectively. This means that the
triplet $\left( U^{\ast },X^{\ast },V^{\ast }\right) $ is observed
if and only if $U^{\ast }\leq X^{\ast }\leq V^{\ast }$, while no
information is
available when $X^{\ast }<U^{\ast }$ or $X^{\ast }>V^{\ast }$.
We assume that the truncation comes from the existence of an observational window of non-random length $\phi$, and therefore $V^{\ast }= U^{\ast}+ \phi$ $(\phi>0)$. In Survival Analysis this model is suitable when the sample reduces to individuals with event dates between two fixed calendar times (e.g. Moreira and de U\~na{-}\'Alvarez \cite{Moreira10}).  Austin and Betensky \cite{Austin2014} termed this type of truncation as 'complete truncation dependence', while Zhu  and  Wang (\cite{Zhu2012}; \cite{Zhu2014}) referred this problem as 'interval sampling'. Efron and Petrosian \cite{Efron99} introduced the NPMLE of $F$ under the assumption of independence between the truncation variables and the target $X^*$. In order to generalize Efron-Petrosian estimator, we assume that
$U^*$ may depend on the variable of interest, and that the dependence structure of $(X^*,U^*)$ is given by a copula function such that (cfr. Nelsen \cite{Nelsen2006})
\[
P(X^{\ast}\leq x, U^{\ast}\leq u)=\mathcal{C}_{\theta}\left(F(x), K(u)\right),
\] where $K(u)=G(u,\infty)$ is the marginal df of $U^*$ and $\mathcal{C}_{\theta}$ a parametric family of copula's, with $\theta$ belonging to a certain euclidean parametric space $\Theta$. With interval sampling, dependent truncation may appear when, for example, the birth date of the process ($U^*$) has influence on the subsequent variable of interest ($X^*$); Austin and Betensky \cite{Austin2014} introduced a test for independence based on a Kendall's Tau in this setting. For example, in the study of transfusion-related AIDS in Section \ref{sec:Section4}, the incubation time $X^*$ is doubly truncated by the time from HIV infection to January 1, 1982 ($U^*$) and the lapse time from HIV infection to the end of study (July 1, 1986) ($V^*$). Hereby we note that several persons in this study were infected a long time ago with the HIV virus without developing AIDS. AIDS was not well known in the early days of the epidemic and, therefore, infected people may have gone unnoticed for some time; this could result in a positive dependence between $X^*$ and $U^*$. This dependence is confirmed in Section ~\ref{subsection41}, where the copula parameter is found to be significantly different to the value corresponding to independence. The situation is the opposite for the Immune Reconstitution Inammatory Syndrome (IRIS) data analyzed in Section ~\ref{Subsection42}, where the independence assumption between the truncation variables and the target can be accepted. \\
For one-sided truncation, Chaieb et al. \cite{Chaieb06} proposed a semiparametric copula model to assess the degree of dependence between the variable of interest and truncation variable. In the same setting of one-sided truncation, Emura et al. \cite{Emura2011} and Emura and Wang \cite{EMURA2012} considered estimators based on conditional likelihood and nonparametric likelihood, respectively. The referred author Emura \cite{Emura2015} revisited the estimation approach in Chaieb  et  al. \cite{Chaieb06} and proposed a different algorithm to solve their estimation function. For the best of our knowledge no similar approach has been proposed so far for doubly truncated data. This paper presents new statistical methods for modelling a possible dependency between $X^{\ast}$ and $(U^{\ast}, U^{\ast}+ \phi)$ when only triplets such that $U^{\ast }\leq X^{\ast }\leq U^{\ast }+\phi$ are observed. \\

$\ $Let $%
\left( U_{i},X_{i},V_{i}\right) $, $i=1,...,n$, denote the sampling
information, these are iid data with the distribution of
$\left( U^{\ast },X^{\ast },V^{\ast }\right) $ conditionally on $U^{\ast }\leq
X^{\ast }\leq V^{\ast }$.
Under the given model where $V=U+\phi$ and $X$ and $U$ are linked through the copula function, the full likelihood of the data can be written as depending on the $X$-masses $f_i$, the $U$-masses $k_i$ and the weights $W_{ij}=\mathcal{C}_{\theta}^{(1,1)}(F_i, K_j)$, with $F_i=\displaystyle \sum_{m=1}^n f_mI_{[X_m \leq X_i]}$ and $K_i=\displaystyle \sum_{m=1}^n k_mI_{[U_m \leq U_i]}$, where $\mathcal{C}_{\theta}^{(1,1)}$ denotes the density of the copula family, as:
 \begin{eqnarray} {L}(\theta, f, k)=\displaystyle \prod_{i=1}^n \frac{W_{ii}f_ik_i}{\displaystyle \sum_{j=1}^{n}\displaystyle \sum_{m=1}^n W_{jm}  f_jk_m J_{mj}}. \end{eqnarray} \label{eq0} Further details are provided in Appendix A.1.\\
 For independent truncation, we have $\mathcal{C}_{\theta}^{(1,1)}=1$ and the likelihood (\ref{eq0}) reduces to that in \cite{Efron99}. When $X^*$ and $U^*$ are dependent, the weights $W_{ij}$ introduce a suitable correction of the Efron-Petrosian NPMLE. The goal is the estimation of the $r+2n$ parameters $\theta$, $f_i$ and $k_i$, $i=1,\ldots,n$, where $r$ denotes the dimension of $\Theta$. Then, the NPMLE's of $F(x)$ and $K(u)$ under truncation are simply obtained as $\widehat F(x)= \displaystyle \sum_{i=1}^n \widehat f_i I_{[X_i\leq x]}$ and $\widehat K(u)= \displaystyle \sum_{i=1}^n \widehat k_i I_{[U_i\leq u]}$,  where $\hat f_i$ and $\hat k_i$ are the maximizers for equation (\ref{eq0}).\\
 For one-sided truncation approaches other than the copula-based have been suggested in order to cope with dependence, including transformation approaches and frailty models \cite{Chiou2019}. Frailty models operate under the assumption of conditional independence between the target and the truncation variables given the frailty variable; they are closely related to copula models in the sense that both the frailty parameter and the nonparametric margins must be estimated. On the other hand, the transformation approach assumes the existence of a latent, independent truncation variable that would have been attached to the target in the absence of dependence; full theoretical developments for this approach as well as extensions beyond one-sided truncation are missing however. Copula-based estimators  have been successfully used in many instances with incomplete data; in this paper we show that copulas are convenient to model for dependence under double truncation too.\\
This paper is organized as follows. In Section ~\ref{sec:Section2} two different algorithms to estimate the parameter  $\theta$ and the distributions $F$ and $K$ are introduced. The finite sample behaviour of the estimators is investigated through simulations in Section ~\ref{sec:Section3}. An application to the analysis of AIDS incubation times and Immune Reconstitution Inflammatory Syndrome (IRIS) diagnosis times are provided in Section ~\ref{sec:Section4}, while the conclusions are deferred to Section ~\ref{sec:Section5}. Technical details are provided in the Appendices A.1 - A.3.

\section{The estimators}
\label{sec:Section2}

The algorithm that we propose optimizes the global likelihood in pieces, i.e, in which $\theta$, $K$ and $F$ are optimized iteratively. For $K$ and $F$, we set up the score equations which are solved in two different ways:  the simple algorithm which have the same philosophy as the iterative method of Jacobi to solve a set of linear equations. For the full algorithm, there is a similar link to the numerical method of Gauss-Seidel.\\
First, we introduce a simple algorithm to estimate the parameters. Here we assume for the moment that the weights $W_{ij}$ are free of $f$ and $k$. Then, by differentiating the loglikelihood with respect to the $f_m$'s and $k_m$'s we obtain the following simple score equations:\\

\begin{equation}
\frac{\partial log L}{\partial f_m}=0 \Leftrightarrow f_m=\left[\displaystyle \sum _{i=1}^n \frac 1{K_i^w}\right]^{-1}\frac1{K_m^w}, \enspace m=1, \ldots, n
\label{eq1}
\end{equation}

\noindent with $K_i^w=\displaystyle \sum _{j=1}^mW_{ij}k_jJ_{ji}$, and \\

\begin{equation}
\frac{\partial log L}{\partial k_m}=0 \Leftrightarrow k_m=\left[\displaystyle \sum _{i=1}^n \frac 1{F_i^w}\right]^{-1}\frac1{F_m^w},\enspace m=1, \ldots, n
\label{eq2}
\end{equation}

\noindent with $F_i^w=\displaystyle \sum _{j=1}^mW_{ij}f_jJ_{ij}$. Equations (\ref{eq1}) and (\ref{eq2}) can be used to introduce the
following iterative simple algorithm.

\begin{enumerate}
\item [Step 0] Take  the Efron-Petrosian NPMLE for independent truncation $ f^{(0)}=(f_1^{EP},...,f_n^{EP})$, $ k^{(0)}=(k_1^{EP},...,k_2^{EP})$ as initial solution for $f$ and $k$, and compute
\[
\theta^{(0)}=\mathnormal {\mbox{argmax}_{\theta}}L^{(0)}(\theta),
\] where,
\begin{eqnarray*}
L^{(0)}(\theta)=\displaystyle \prod_{i=1}^n \frac{\mathcal {C}_{\theta}^{(1,1)}\left(F_i^{(0)}, K_i^{(0)}\right)f_i^{(0)}k_i^{(0)}}{\displaystyle \sum_{j=1}^{n}\displaystyle \sum_{m=1}^n {\mathcal {C}_{\theta}^{(1,1)}\left(F_j^{(0)}, K_m^{(0)}\right)  f_j^{(0)}k_m^{(0)} J_{mj}}},
\end{eqnarray*}
and where $F_i^{(0)}=\displaystyle \sum_{m=1}^n f_m^{(0)}I_{[X_m \leq X_i]}$ and $K_i^{(0)}=\displaystyle \sum_{m=1}^n k_m^{(0)}I_{[U_m \leq U_i]}$

\item [Step 1] Use (\ref{eq2}) to improve $k^{(0)}$:
\begin{eqnarray*}
k_m^{(1)}=\left[\displaystyle \sum _{i=1}^n \frac 1{F_i^{w_0,0}}\right]^{-1}\frac 1{F_m^{w_0,0}}, \enspace m=1, \ldots, n
\end{eqnarray*} where $w_0=\{W_{ij}^{(0)}: 1 \leq i, j \leq n\}$, $W_{ij}^{(0)}= \mathcal{C}_{\theta^{(0)}}^{(1,1)}\left(F_i^{(0)}, K_j^{(0)}\right)$
and $F_i^{w_0,0}=\displaystyle \sum_{j=1}^n W_{ij}^{(0)}f_j^{(0)}J_{ij}$

\item [Step 2] Use (\ref{eq1}) to improve $f^{(0)}$:
\begin{eqnarray*}
f_m^{(1)}=\left[\displaystyle \sum _{i=1}^n \frac 1{K_i^{w_0,1}}\right]^{-1}\frac 1{K_m^{w_0,1}}, \enspace m=1, \ldots, n
\end{eqnarray*} where  $K_i^{w_0,1}=\displaystyle \sum_{j=1}^n W_{ij}^{(0)}k_j^{(1)}J_{ji}$

\item [Step 3] Improve $\theta^{(0)}$ by taking
\[
\theta^{(1)}=\mathnormal{\mbox{argmax}_{\theta}}L^{(1)}(\theta),
\] where
\begin{eqnarray*}
L^{(1)}(\theta)=\displaystyle \prod_{i=1}^n \frac{\mathcal {C}_{\theta}^{(1,1)}\left(F_i^{(1)}, K_i^{(1)}\right)f_i^{(1)}k_i^{(1)}}{\displaystyle \sum_{j=1}^{n}\displaystyle \sum_{m=1}^n {\mathcal {C}_{\theta}^{(1,1)}\left(F_j^{(1)}, K_m^{(1)}\right)  f_j^{(1)}k_m^{(1)} J_{mj}}},
\end{eqnarray*}
and where $F_i^{(1)}=\displaystyle \sum_{m=1}^n f_m^{(1)}I_{[X_m \leq X_i]}$, $K_i^{(1)}=\displaystyle \sum_{m=1}^n k_m^{(1)}I_{[U_m \leq U_i]}$\\

\item [Step 4] Repeat steps $(1)-(3)$ until convergence.
\end{enumerate}

That is, algorithm Step 0-Step 4 fits the copula function by starting with the Efron-Petrosian NPMLE estimator under independent truncation. Then, it improves first $k$ and then $f$ by using the simple score equations (\ref{eq2}) and (\ref{eq1}); and, finally, it updates $\theta$ by maximizing the loglikelihood (based on the improved $k$ and $f$) with respect to the copula parameter. This procedure is repeated until a stable solution is reached. As convergence criterion, we have used $\displaystyle \max_{1 \leq i \leq n}|f_i^{q-1}-f_i^q|\leq 1e-06$ and $\displaystyle \max_{1 \leq j \leq n}|k_j^{q-1}-k_j^q|\leq 1e-06$ and $\displaystyle \max|\theta^{q-1}-\theta^q|\leq 1e-06$. Then, the NPMLE's $\widehat F(x)$ and $\widehat K(u)$ are constructed from the q-th solution $f_i^q$, $k_j^q$ and $\theta^q$. \\

A second algorithm to estimate the different parameters is called the full algorithm and is obtained if one differentiates the loglikelihood with respect to $f$ and $k$ by taking the dependence of the $W_{ij}$'s on these parameters into account. Then, the substitutes for equations (\ref{eq2}) and (\ref{eq1}) are (see the Appendix A.1 for details):\\

\begin{eqnarray}
&&\frac{\partial log L}{\partial f_m}=0 \Leftrightarrow \nonumber\\
 &&f_m=\left[\displaystyle \sum _{i=1}^n \frac 1{{nA_i}+n K_i^w{- \alpha B_i}}\right]^{-1}\frac1{{nA_m}+nK_m^w{-\alpha B_m}}, \enspace m=1, \ldots, n
\label{eq1full}
\end{eqnarray}

\noindent with
$A_m=\displaystyle \sum_{i=1}^n \displaystyle \sum _{j=1}^n W_{ij}^{(2,1)}f_ik_j J_{ji}I_{[X_m \leq X_i]}$,
$B_m=\displaystyle \sum_{i=1}^n\frac{W_{ii}^{(2,1)}I_{[X_m \leq X_i]}}{W_{ii}^{(1,1)}}$ and\\
$
\alpha= \displaystyle \sum_{j=1}^{n} \displaystyle \sum_{m=1}^n W_{jm}^{(1,1)}  f_jk_m J_{mj}
$, and

\begin{eqnarray}
&&\frac{\partial log L}{\partial k_m}=0 \Leftrightarrow \nonumber\\
 &&k_m=\left[\displaystyle \sum _{i=1}^n \frac 1{{nC_i}+n F_i^w{- \alpha D_i}}\right]^{-1}\frac1{{nC_m}+nF_m^w{-\alpha D_m}}, m=1, \ldots, n
\label{eq2full}
\end{eqnarray}

with
$C_m=\displaystyle \sum_{i=1}^n \displaystyle \sum _{j=1}^n W_{ij}^{(1,2)}f_ik_j J_{ji}I_{[U_m \leq U_i]}$ and $D_m=\displaystyle \sum_{i=1}^n\frac{W_{ii}^{(1,2)}I_{[U_m \leq U_i]}}{W_{ii}^{(1,1)}}$.\\

In (\ref{eq1full}) and (~\ref{eq2full}) we use the notation $W_{ij}^{(l,m)}$, $1\leq l,m\leq 2$, for $\mathcal C_{\theta}^{(l,m)}(F_i,K_j)$, where  $\mathcal C_{\theta}^{(l,m)}(u,v)=\frac {\partial^{l+m}}{\partial u^l\partial v^m} \mathcal C_{\theta}(u,v)$. Note that $W_{ij}=W_{ij}^{(u,v)}$ with this notation.

The 'full' algorithm we propose is defined following Steps 0-4 above, but using these two equations  (~\ref{eq1full}) and (~\ref{eq2full}) in the place of  (\ref{eq2}) and (\ref{eq1}). Note that moving from the simple algorithm to this full algorithm implies changing the way in which $k$ and $f$ are improved, while the updating of $\theta$ (Step 0) remains the same.\\

In Section~\ref{sec:Section3} we investigate through simulations the performance of these two algorithms for several copula functions and marginal models. Interestingly, it is seen that the simple algorithm is accurate enough for practical purposes, while giving a more efficient solution in terms of computational speed. For the final implementation we multiply $A_m$, $B_m$, $C_m$ and $D_m$ by $n/{n+1}$; this is equivalent to replace each $W_{ij}=\mathcal {C}_{\theta}^{(1,1)}\left(F_i, K_j\right)$ by $W_{ij}^*= \mathcal {C}_{\theta}^{(1,1)}\left(\frac n {n+1}F_i, \frac n {n+1} K_j\right)$, which avoids problems at the upper-right corner of the copulas function.

Since the truncation interval $(U^*,V^*)$ prevents us from always observing the variable of interest $X^*$, we are not able to fully see the dependence structure between $(X^*,U^*)$ which was expressed using a copula function $\mathcal{C}$. Hence, it is not possible to estimate the association copula function and the marginal distributions without introducing extra assumptions. In Appendix A.2 we formally study the identifiability of the copula model, following ideas similar to those in Ding \cite{Ding2012} for dependent left-truncation. The copula functions used in the simulations (Section~\ref{sec:Section3}) and real data analyses (Section~\ref{sec:Section4}) fulfill the given identifiability condition.

In practice, it is important to report standard errors to know the accuracy of a given estimator for the triplet $(\theta,F,G)$. To this end, we propose to use a bootstrap algorithm based on the fitted chosen copula. To be specific, let $(T_1,T_2)$ be a pair of $U(0,1)$ random variables following the fitted copula $\mathcal{C}_{\widehat \theta}$. Let $U^*=\widehat K^{-1}(T_1)$ and $X^*=\widehat F^{-1}(T_2)$ where $\widehat F$ and $\widehat K(.)=\widehat G(.,\infty)$ are the estimators based on the simple or the full algorithm, and $\widehat F^{-1}$ and $\widehat K^{-1}$ are their respective quantile functions. Reject the pair $(U^*,X^*)$ if $U^*\leq X^* \leq U^*+\phi$ is violated. Form a resample of $n$ data following this scheme, and repeat up to forming $B$ resamples. Then, the bootstrap standard error of $\widehat \theta$, $\widehat F$ or $\widehat G$ is defined as the standard deviation of these estimators along the $B$ resamples. In Section ~\ref{sec:Section3} we include some simulation results for this method when the goal is the estimation of the standard error of $\widehat \theta$; these results suggest that the copula-based bootstrap performs well.

\section{Simulations}
\label{sec:Section3}

In this section we investigate the finite sample performance of the algorithms proposed in Section \ref{sec:Section2} through simulations. We simulate the scenario $X^* \sim U(0,1)$, $U^* \sim U(-0.6, 0.4)$ and then we take $V^*=U^* + \phi$, with  $\phi=1.5$. Note that, in this way, the df of $X^*$ is identifiable, because the lower (resp. upper) limit of the support of $U^*$ (resp. $V^*$) is smaller than the lower (resp. upper) limit of the support of $X^*$ \cite{Woodroofe85}. We consider three different copula families: the Farlie-Gumbel-Morgentein (FGM) copula (Case 1), the Frank copula (Case 2) and the Clayton copula (Case 3). \\
In Case 1 the variables $X^*$ and $U^*$ follow a FGM copula family with parameter $\theta$, that is, $\mathcal{C}_{\theta}(u_1,u_2)=u_1u_2+\theta u_1u_2 (1-u_1)(1-u_2), \theta \in [-1,1]$. The Kendall's Tau ($\tau_{\theta}$) corresponding to this copula is $\tau_{\theta}=\frac29\theta$. We consider the cases $\theta= -1, -0.5, 1$ reporting association levels between $X^*$ and $U^*$ equal to $-0.2,-0.1$ and $0.2$ (Models 1.1-1.3 respectively). Specifically, the simulation algorithm is as follows (cfr. Exercise 3.23 in Nelsen \cite{Nelsen2006}):
  \begin{enumerate}
\item [Step 1] Generate two independent uniform $(0,1)$ variables $X^*$ and $T$;
\item [Step 2] Set $a=1+ \theta(1-2X^*)$ and $b=\sqrt{(a^2-4(a-1)T)}$;
\item [Step 3] Set $U^*=\frac{a-b}{2(a-1)}$;
\item [Step 4] Update $U^*$ to be $U^*-0.6$ according to its support $(-0.6,0.4)$;
\item [Step 5] The desired pair is $(X^*,U^*)$, satisfying the condition $U^*\leq X^*5\leq U^* + \phi$.\\
 \end{enumerate}

In Case 2  the variables $X^*$ and $U^*$ follow a Frank copula family with parameter $\theta$, given by $\mathcal{C}_{\theta}(u_1,u_2)=-\frac{1}\theta \log\left[1+\frac{\left(e^{-\theta u_1}-1\right)\left(e^{-\theta u_2}-1\right)}{e^{-\theta}-1}\right], \theta \in \mathcal{R}\backslash \{0\}$. For this copula the Kendall's Tau is the solution of the equation $\frac{\left[D_1(\theta)-1\right]}{\theta}=\frac{\tau_\theta-1}4$, where $D_1(\alpha)=\frac1\alpha\int_0^\alpha\frac t{e^t-1}dt$ is a Debye function of the first kind. We consider the cases $\theta= -2.1, -1, 1.86, 5.74, 20.9$ corresponding to association levels of $-0.2,-0.1, 0.2, 0.5$ and $0.9$ respectively (Models 2.1-2.5). The simulation algorithm is as follows (cfr. Exercise 4.17 in \cite{Nelsen2006}):\\
  \begin{enumerate}
\item [Step 1] Generate two independent uniform $(0,1)$ variables $T$ and $U^*$;
\item [Step 2] Set $X^*=-(1/\theta)\log(1+(T(\exp(-\theta)-1))/(T+(1-T)\exp(-\theta \times U^*)))$;
\item [Step 3] Update $U^*$ to be $U^*-0.6$ according to its support $(-0.6,0.4)$;
\item [Step 4] The desired pair is $(X^*,U^*)$, satisfying the condition $U^*\leq X^* \leq U^* + \phi$. \\
 \end{enumerate}

In Case 3  the variables $X^*$ and $U^*$ follow a  Clayton copula family with generator $\psi_{\theta}(t)=\theta^{-1}(t^{-\theta}-1)$, $\theta>0$, i.e., $\mathcal{C}_{\theta}(u_1,u_2)=\left(u_1^{-\theta}+u_2^{-\theta}-1\right)^{-1/\theta} , \theta \in (0,\infty)$. This copula implies a  Kendall's Tau $\tau_{\theta}=\frac \theta{\theta+2}$. We consider the cases $\theta= 0.5, 2, 18$ corresponding respectively to association levels of $0.2,0.5$ and $0.9$ (Models 3.1-3.3). The simulation algorithm is as follows (cfr. Exercise 4.17 in Nelsen \cite{Nelsen2006}):
  \begin{enumerate}
\item [Step 1] Generate independent random variables $Y_1$, $Y_2$ $\sim Exp(1)$;
\item [Step 2] Independently generate $Z_0 \sim \Gamma(1/\theta,1/\theta)$, and compute $U^*=(1+\theta Y_2/Z_0)^{(-1/\theta)}$;
\item [Step 3] Finally compute $X^*=(1+\theta Y_1/Z_0)^{(-1/\theta)}$;
\item [Step 4] Update $U^*$ to be $U^*-0.6$ according to its support $(-0.6,0.4)$;
\item [Step 5] The desired pair is $(X^*,U^*)$, satisfying the condition $U^*\leq X^* \leq U^* + \phi$.\\
 \end{enumerate}

The values of $\theta$  for the several copulas correspond to the same association levels (Kendall's Tau). This will be interesting when interpreting the simulation results.
The simulated scenarios result in different truncations proportions according to the different copula families and parameter values $(\theta)$ considered. For instance, in Case 1, the proportion of truncation ranges from 4\% (Model 1.3) to 13\% (Model 1.1); in Case 2, from 1\% (Model 2.4) to 13\% (Model 2.1); and in Case 3, from 1\% (Model 3.1) to 8\% (Model 3.3). \\

\begin{figure*}
\centering
%\begin{tabular}{cc}
\includegraphics[width=0.45\textwidth]{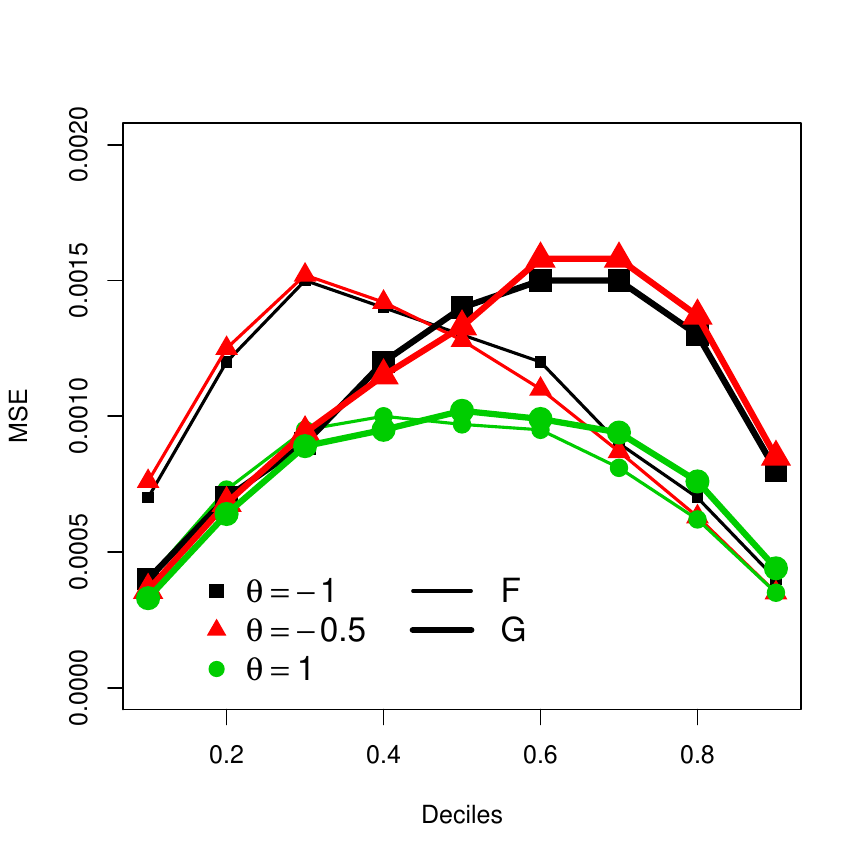} ~
\includegraphics[width=0.45\textwidth]{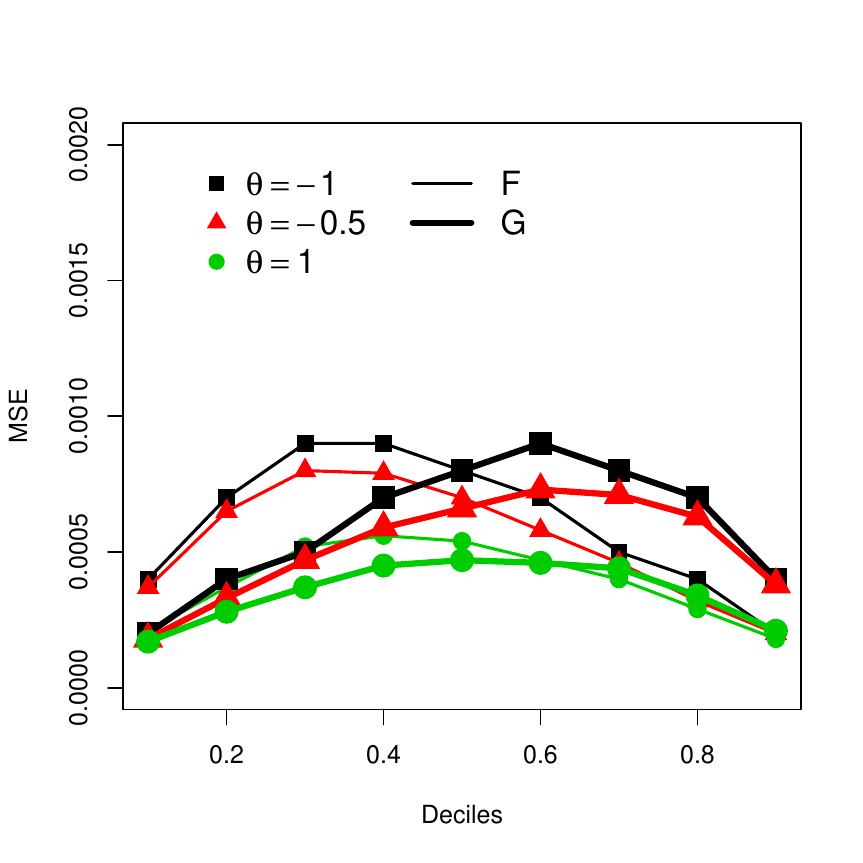} \\
\includegraphics[width=0.45\textwidth]{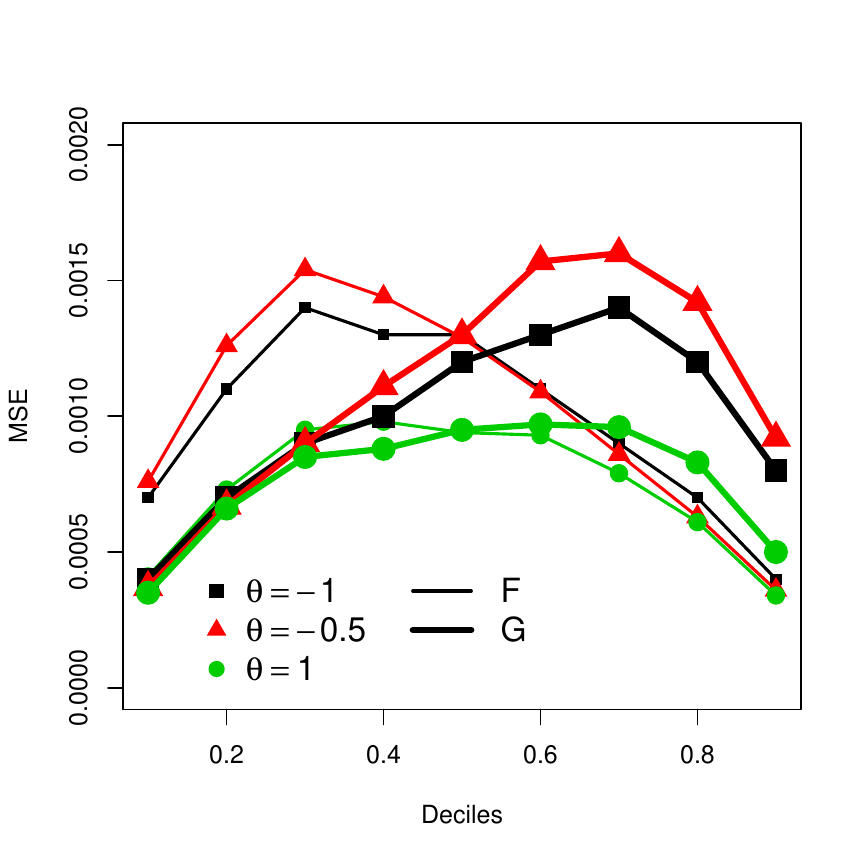} ~
\includegraphics[width=0.45\textwidth]{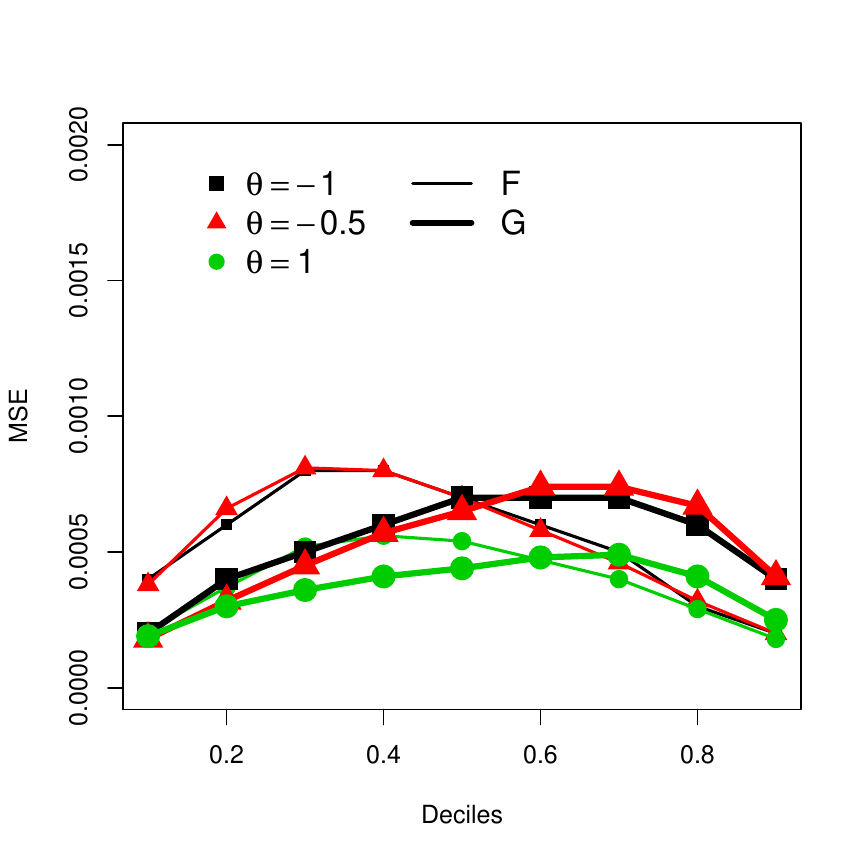}\\
%\end{tabular}
\caption{ MSE's of the proposed estimators $\widehat F$ and $\widehat G$, in each decile, for FGM copula. $n=250$ (left) $n=500$ (right), for simple and full algorithms (from top to bottom).  FGM copula.} \label{Figure11}
\end{figure*}

\begin{figure}[!h]
\begin{minipage}[b]{0.45\linewidth}
\includegraphics[width=\textwidth]{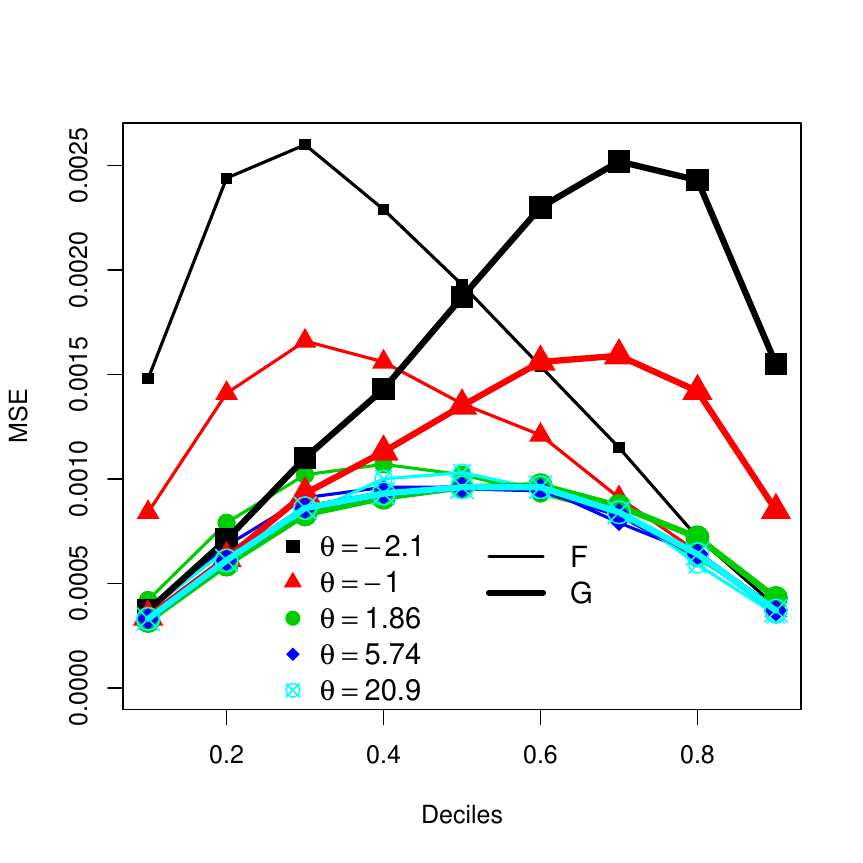}
\end{minipage} \hfill
\begin{minipage}[b]{0.45\linewidth}
\includegraphics[width=\textwidth]{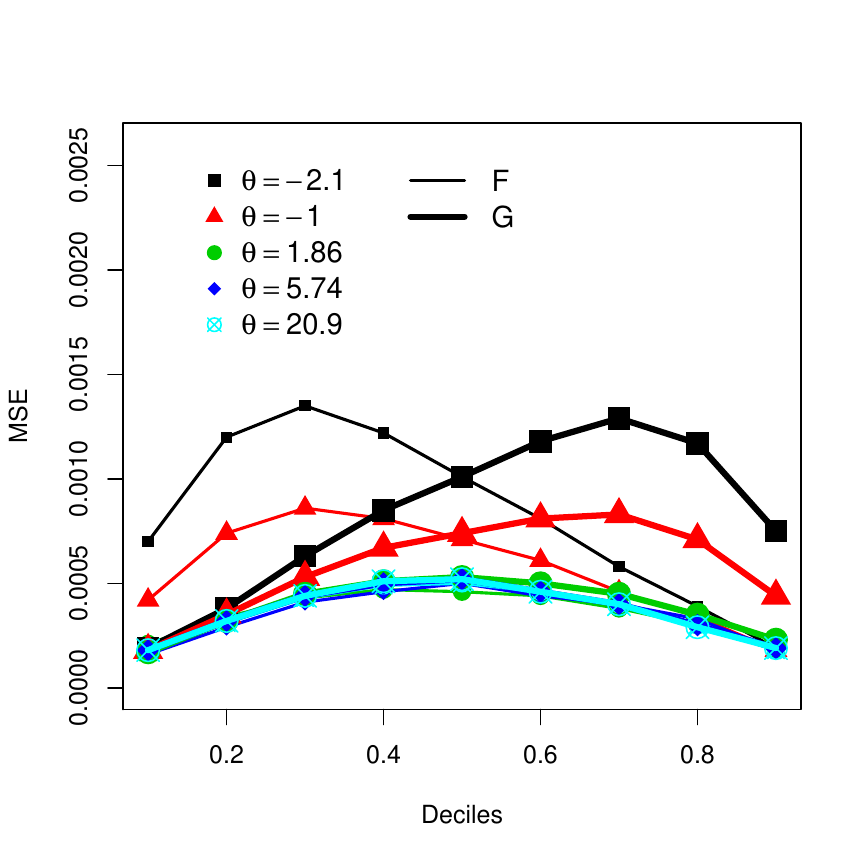}
\end{minipage} \hfill
\vspace{-0.2cm}
\caption{ MSE's of the proposed estimators $\widehat F$ and $\widehat G$, in each decile, for Frank copula. $n=250$ (left) $n=500$ (right).  Frank copula.} \label{Figure33}
\end{figure}

\begin{figure}[!h]
\begin{minipage}[b]{0.45\linewidth}
\includegraphics[width=\textwidth]{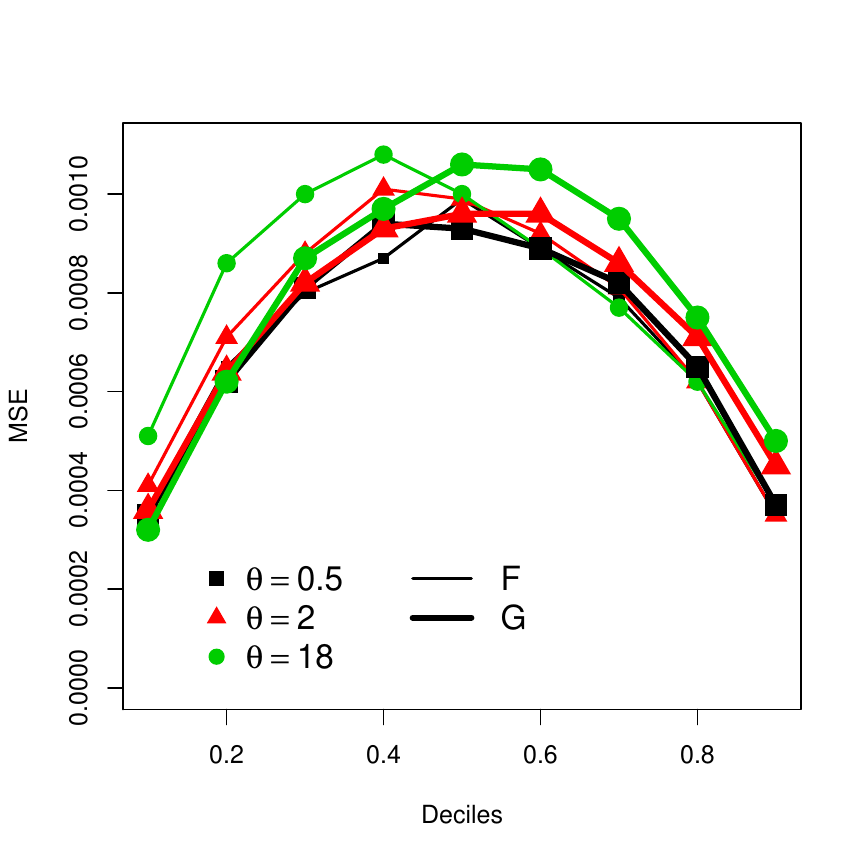}
\end{minipage} \hfill
\begin{minipage}[b]{0.45\linewidth}
\includegraphics[width=\textwidth]{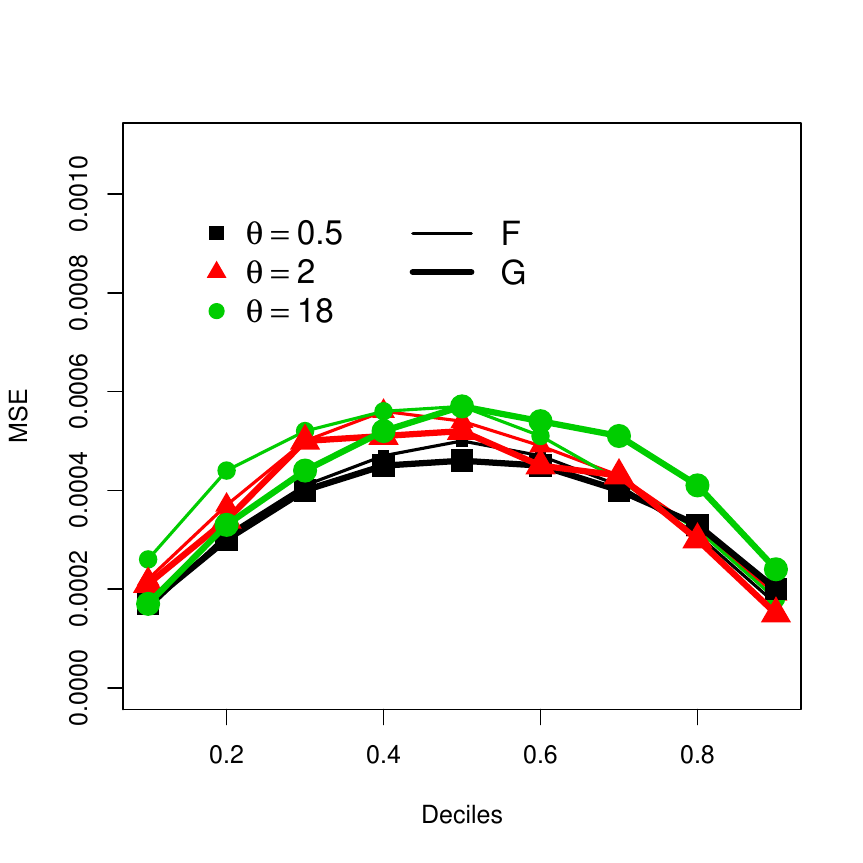}
\end{minipage} \hfill
\vspace{-0.2cm}
\caption{ MSE's of the proposed estimators $\widehat F$ and $\widehat G$, in each decile, for Clayton copula. $n=250$ (left) $n=500$ (right). Clayton copula.} \label{Figure22}
\end{figure}

In Figures~\ref{Figure11} to~\ref{Figure22} we report the MSE of the proposed estimators ($\widehat F$ and $\widehat G$) for each $\theta$ and for the several copulas, computed along 1000 Monte Carlo trials of size $n=250$ and $n=500$, at the deciles of the distribution of $X^*$. We performed simulations for lower sample sizes ($n=50, 100$) too, reporting similar results (see Appendix A.3). The MSEs decrease when increasing the sample size thus suggesting the consistency of the proposed methods. In Figure~\ref{Figure11} (Models 1.1 to 1.3, FGM copula) we report the results of both simple and full algorithms (top from bottom).
In these figure we see that, in general, the simple algorithm provides MSE's slightly larger than those of the full algorithm. Since the full algorithm is computationally heavier see Table~\ref{Table42}, we have evaluated the relative increase of the MSE when moving from the full to the simple algorithm ($RMSE=(MSE(\mbox{simple})-MSE(\mbox{full}))/MSE(\mbox{full})$) for the four sample sizes $n=50, 100, 250, 500$ and all the simulated scenarios. For each copula family we computed the quartiles of the RMSE throughout the several values of $\theta$, the various sample sizes and the two target curves ($F$ and $G$), see Table~\ref{RMSE}. In this table we see that the median increase is only of 1.19\%, 0.65\%, or 0\% depending on the copula (FGM, Frank and Clayton resp.). Besides, by looking at the first quartile of $RMSE$ we see that the simple method is doing it better than the full method at least 25\% of the times. On the other hand, the third quartiles reveal that 75\% of the times the RMSE is below 5\%, 3.8\% or 2.43\% depending again on the copula. Models for which the full algorithm reports the best relative performance are those with large negative association between the variable of interest and the truncation variables, particularly when estimating $G$. Overall, the simple algorithm seems to be the best option according to its good relative performance and computational speed. This is why we only display the results corresponding to the simple algorithm for Frank and Clayton copulas.

%The Tables \ref{Table7} to \ref{Table11} the results are for Case 2 (Models 2.1 to 2.5) and Tables \ref{Table12} to \ref{Table14} report the results of Case 3 (Models 3.1 to 3.3).

\begin{table}[ht]
\caption {The quartiles and mean of the overall RMSE's throughout the several values of $\theta$, the various sample sizes and the two target curves $F$ and $G$.}
\label{RMSE}
\begin{center}
\begin{tabular}{ccccc}
  \hline
 & $Q_1$ &$Q_2$& $Q_3$ & Mean \\
 \hline
FGM &  -0.0045&0.0119&0.0500&0.0221\\
Frank & -0.0166&0.0065&0.0280&0.0026\\
Clayton&  -0.0248 & 	0.0000 &0.0243 & 0.0010 \\
   \hline
\end{tabular}
\end{center}
\end{table}

In Table \ref{Theta} we display the bias and standard deviation of the estimator $\widehat \theta$ obtained from the simple algorithm along the 1,000 trials, for each copula function and sample sizes $n=250, 500$. As expected, it is seen that the bias and the standard deviation decrease when increasing the sample size. The bias and the standard deviation get larger as the association degree increases, although an exception to this is found for the standard deviation and FGM copula. In Appendix A.3, Table~\ref{Thetasmall}, additional simulation results for $n=50$ and $n=100$ can be found.

%As we approximate the independent case $\theta=0$, the bias and the standard deviation decrease.
%\textcolor{red}{ incluir tabla con resultados del estimador de $\theta$}

\begin{table}[ht]
\caption {The bias and the standard deviation of the estimator $\widehat \theta$ obtained from the simple algorithm along the 1,000 trials, for each copula function and sample sizes $n=250, 500$.}
\begin{center}
\begin{tabular}{ccccc}
  \hline
Copula& n& $\theta$& Bias($\widehat \theta$)& sd($\widehat \theta$)\\
 \hline
& & -1& 0.0712&0.1133\\
&250&-0.5&-0.0122&0.2307\\
&&1&-0.0851&0.1293\\
\cline{2-5}
FGM& & -1& 0.0505&0.0817\\
&500 & -0.5& -0.0020&0.1634\\
& & 1& -0.0574&0.0915\\
\hline
\hline
& & -2.1& -0.0844&0.6568\\
&&-1&-0.0482&0.5358\\
&250&1.86&0.0086&0.4535\\
&&5.74&0.0327&0.5453\\
&&20.9&-0.1879&1.3279\\
\cline{2-5}
Frank & & -2.1& -0.0125&0.4609\\
& & -1& -0.0063&0.3904\\
& 500& 1.86&0.0076&0.3353\\
& & 5.74&0.0130&0.3895\\
& & 20.9&-0.1145&0.9041\\
\hline
\hline
& & 0.5& 0.0564&0.0725\\
&250&2&-0.0723&0.1338\\
&&18&-0.0852&0.2523\\
\cline{2-5}
Clayton& & 0.5& 0.0412&0.0548\\
&500 & 2& 0.0523&0.0929\\
& & 18& 0.0684&0.1786\\
\hline
\hline
\end{tabular}
\label{Theta}\end{center}
\end{table}

\begin{figure*}
\centering
%\begin{tabular}{ccc}
\includegraphics[width=0.4\textwidth]{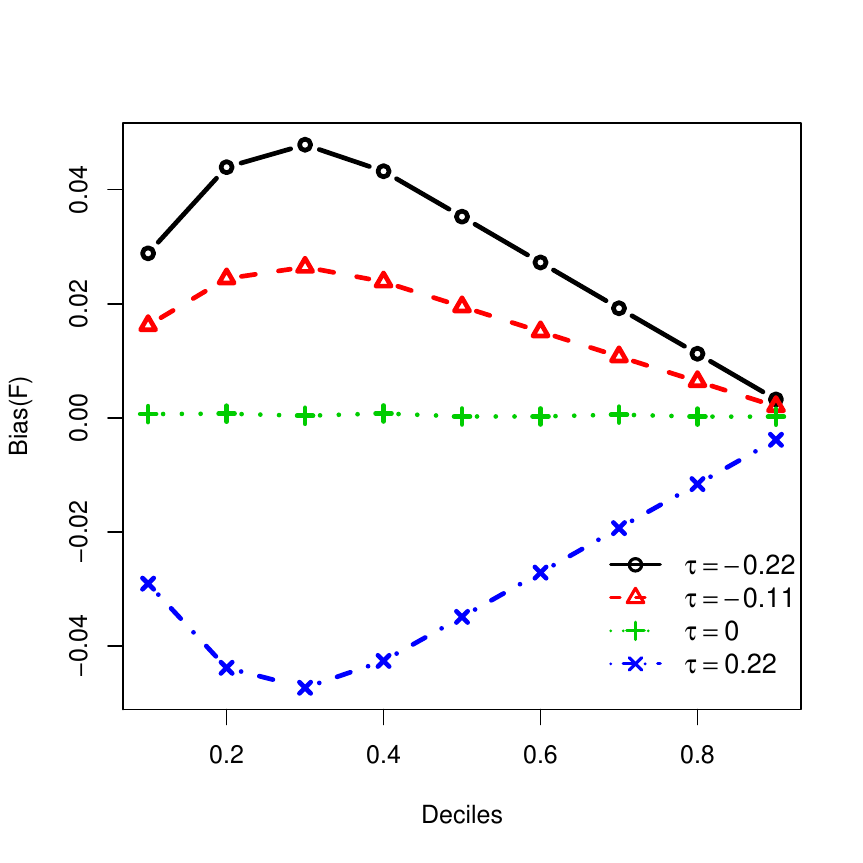} ~
\includegraphics[width=0.4\textwidth]{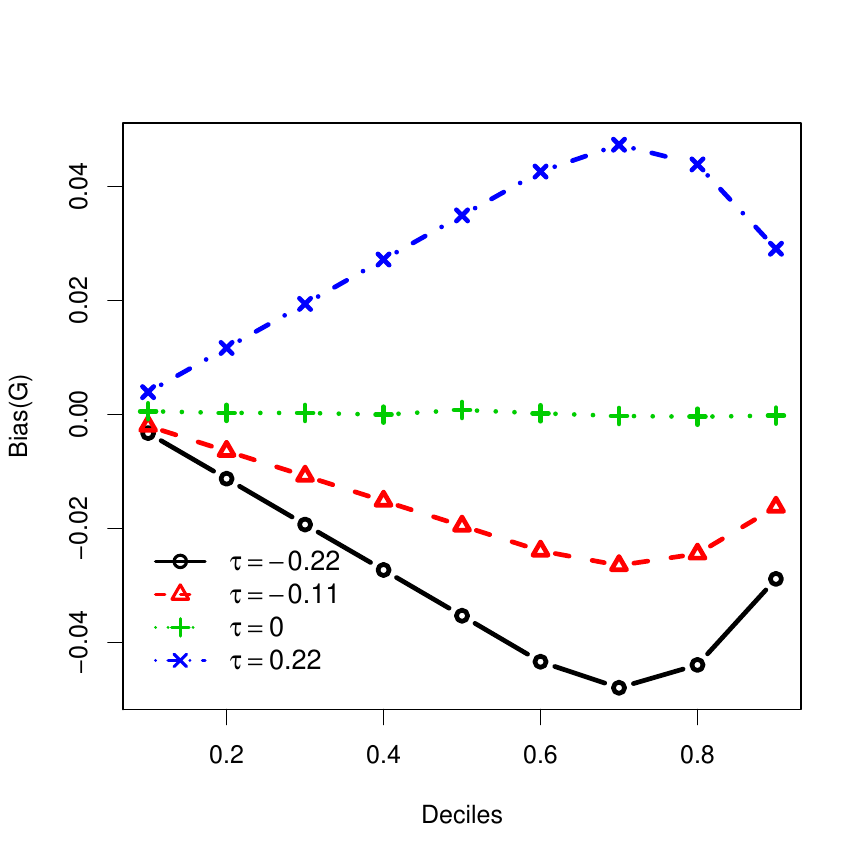} \\
\includegraphics[width=0.4\textwidth]{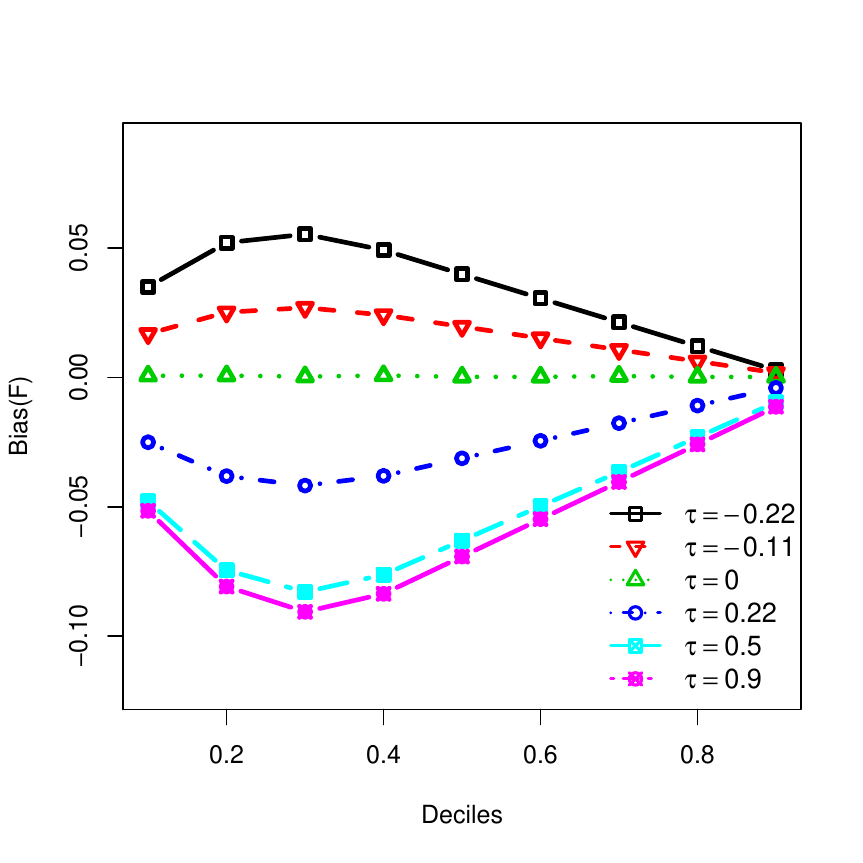}~
\includegraphics[width=0.4\textwidth]{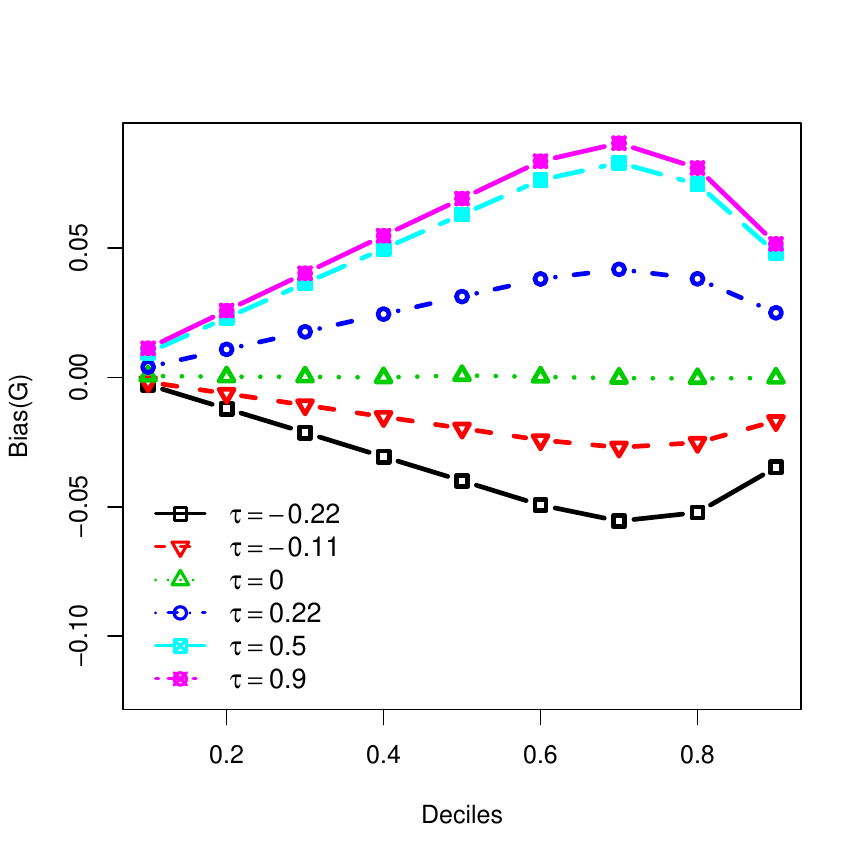}\\
\includegraphics[width=0.4\textwidth]{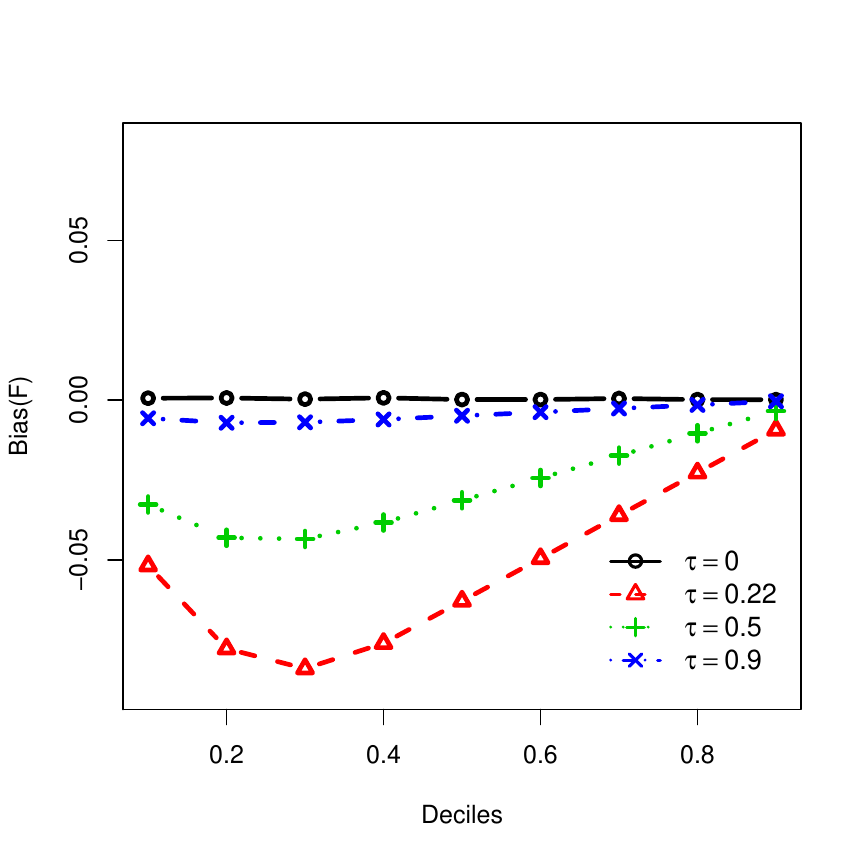} ~
\includegraphics[width=0.4\textwidth]{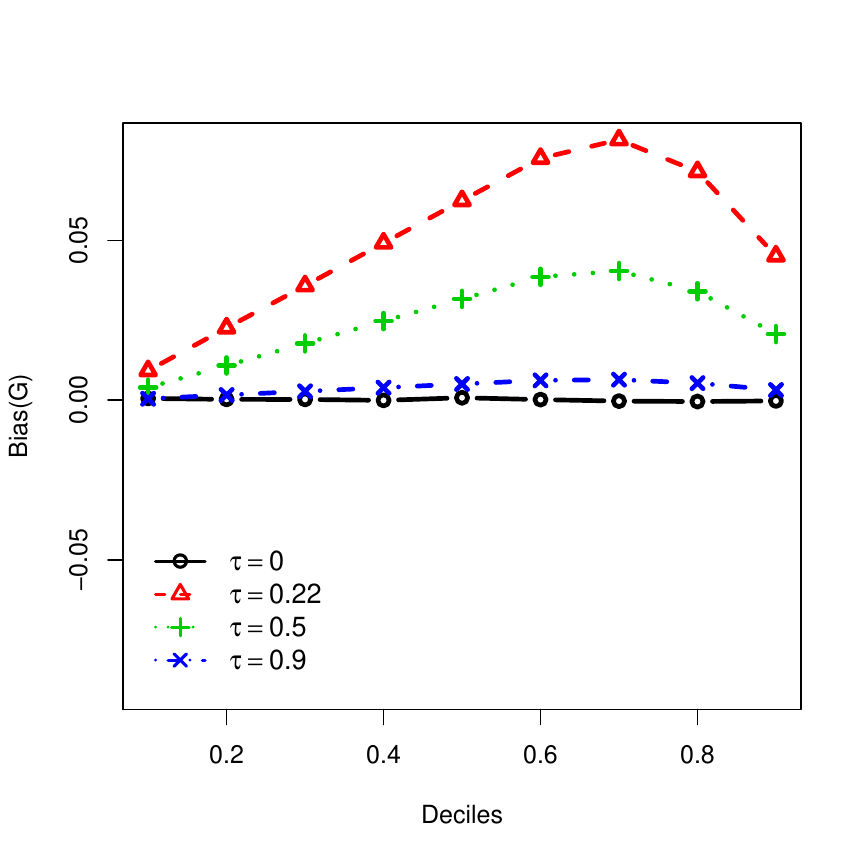} \\
\caption{ Bias of the NMPLE proposed by Shen, in each decile, and each functions $F$ (left) and $G$ (right), for FGM, Frank and Clayton copulas (from top to bottom), with sample size
$n=500$ and different $\tau$'s.} \label{Figure1}
\end{figure*}

We have computed the bias and variance of the NPMLE proposed by Shen \cite{Shen08} for the functions $F$ and $G$, which ignores the possible dependence between $X^*$ and $U^*$. While the variance of the NPMLE and that of the copula-based estimator are of the same order (results not shown), the bias of the NPMLE can be two orders of magnitude larger than that corresponding to the proposed estimator. This can be seen from Figure \ref{Figure1}, in which the bias of the NPMLE for the three copulas under several dependence degrees is depicted for $n=500$ (the case $n=250$ reported similar results). As expected, this bias becomes more visible as the association level grows. For instance, in Case 1, the bias of the NPMLE of $F$ when $\theta=1$ is approximately 1.8 times that corresponding to $\theta=-0.5$ (Figure \ref{Figure1}, top left panel); similar results hold for $\widehat G$ (Figure \ref{Figure1}, top right panel). In Case 2, the bias of the NPMLE of $F$ when $\theta=20.9$ is approximately 2.4 times that corresponding to $\theta=1.86$. \\

%\subsection{Standard error estimation}

As mentioned in Section \ref{sec:Section2}, the bootstrap method can be applied to estimate the standard error of both the marginal distributions and the copula parameter. We have evaluated the performance of the copula-based bootstrap method when estimating the standard error of $\widehat \theta$. To this end, we have computed the ratio between the bootstrap standard error and the true standard deviation of $\widehat \theta$ along 500 Monte Carlo trials (the true standard error was approximated by the Monte Carlo standard deviation). In Table \ref{Table42} we report the mean and the standard deviation of this ratio $Q$ along the simulated runs for the three copula functions with and sample sizes $n=50$ and $n=250$, in order to investigate the performance of the bootstrap when the sampling information is scarce to moderate. From this table it is seen that the bootstrap performs well, giving a more accurate estimation of the standard error of $\widehat \theta$ as the sample size increases.

 \begin{table}[!ht]
 \caption {Mean and standard error of the quotient $Q$.} \label{Table42}
\centering
\begin{tabular}{c c c c }
\hline
n&Copula& mean (Q)& sd(Q)\\
\cline{1-4}
&FGM&0.8804&0.2108\\
50&Clayton&0.9102&0.2019\\
&Frank&1.0828&0.1909\\
\hline
\hline
&FGM&1.0047&0.1135\\
250&Clayton&0.9872&0.1023\\
&Frank&1.0001&0.0900\\
\hline
\end{tabular}
\end{table}

\section{Real data illustrations}
\label{sec:Section4}

\indent For illustration purposes, in this section we consider epidemiological data on transfusion-related Acquired Immune Deficiency Syndrome (AIDS) and Immune Reconstitution Inflammatory Syndrome (IRIS). In both cases the variable of interest is doubly truncated and its independence of the truncation variables may be questionable. For the AIDS data this has been discussed in the Introduction, and in principle a similar situation could hold for the IRIS data described in Section \ref{Subsection42} below.
To assess the possible dependence between the target variable and the truncation variables in our two real data applications we first apply the quasi-independence test based on the conditional Kendall's Tau proposed by \cite{Betensky05} and then we apply our proposed estimator to correct the Efron-Petrosian estimator of $F$ to accommodate such possible dependence.

\subsection{AIDS Blood Transfusion Data}
\label{subsection41}

 The AIDS Blood Transfusion Data are
collected by the Centers for Disease Control (CDC), which is from a
registry data base, a common source of medical data (see
Bilker and Wang \cite{Bilker96}; Kalbfleisch and Lawless \cite{Lawless89}). The variable of interest ($X^*$)
is the induction or incubation time, which is defined as the time
elapsed from Human Immunodeficiency virus (HIV) infection to the
clinical manifestation of full-blown AIDS. The CDC AIDS Blood
Transfusion Data can be viewed as being doubly truncated. The data
were retrospectively ascertained for all transfusion-associated AIDS
cases in which the diagnosis of AIDS occurred prior to the end of
the study, thus leading to right-truncation. Besides, because HIV
was unknown prior to 1982, any cases of transfusion-related AIDS
before this time would not have been properly classified and thus
would have been missed. Thus, in addition to right-truncation, the
observed data are also truncated from the left. See Bilker and Wang \cite{Bilker96}, section 5.2, for further discussions. \\

The data include 494 cases reported to the CDC prior to January 1,
1987, and diagnosed prior to July 1, 1986. Of the 494 cases, 295 had
consistent data, and the infection could be attributed to a single
transfusion or short series of transfusions. Our analyses are
restricted to this subset, which is entirely reported in Kalbfleisch and Lawless
\cite{Lawless89}, Table 1. Values of $U^*$ were obtained by
measuring the time from HIV infection to January 1, 1982; while $V^*$ was defined
as time from HIV infection to the end of study (July 1, 1986). Note that the difference
between $V^*$ and its respective $U^*$ is always 4.5 years. \\

We performed the quasi-independence test of Martin and Betensky \cite{Betensky05}, between the incubation time and the date of HIV infection and we found no evidence to reject quasi-independence assumption ($p =0.15$). However, we proceed with the application of the copula-based extension of the Efron-Petrosian estimator to see if the semiparametric copula model brings evidences on dependence. Specifically, our goal is to correct the Efron-Petrosian estimator of $F$ for the possible dependence between
AIDS incubation time and the date of HIV infection (left truncation variable). In order to assess this dependence, in Table \ref{Table5}
we report the value of $\widehat \theta$ (as well as the corresponding Kendall's Tau $\tau_\theta$) obtained from the two proposed algorithms (full  and simple),
for two copula families (FGM and Frank). The number of iterations needed for each convergence for each algorithm and copula function are included. Bootstrap standard errors and 95\% confidence intervals based on the bootstrap and the normal approximation are reported too. From this Table \ref{Table5} it is seen that (a) the two copulas indicate a positive association between $U^*$ and $X^*$, as it was anticipated in Section \ref{sec:Section1}, and (b) the full algorithm is more computationally demanding. For the FGM copula, we note that the optimal value of $\theta$ is reached at the upper limit of the parameter space of possible $\theta$ values for this copula. This means that the association is in-fact larger than what can be obtained by this copula function. Hence this copula function is not properly suited to look at the association between the incubation time and the truncation time.\\

\begin{figure}[!h]
\begin{minipage}[b]{0.5\linewidth}
\includegraphics[width=\textwidth]{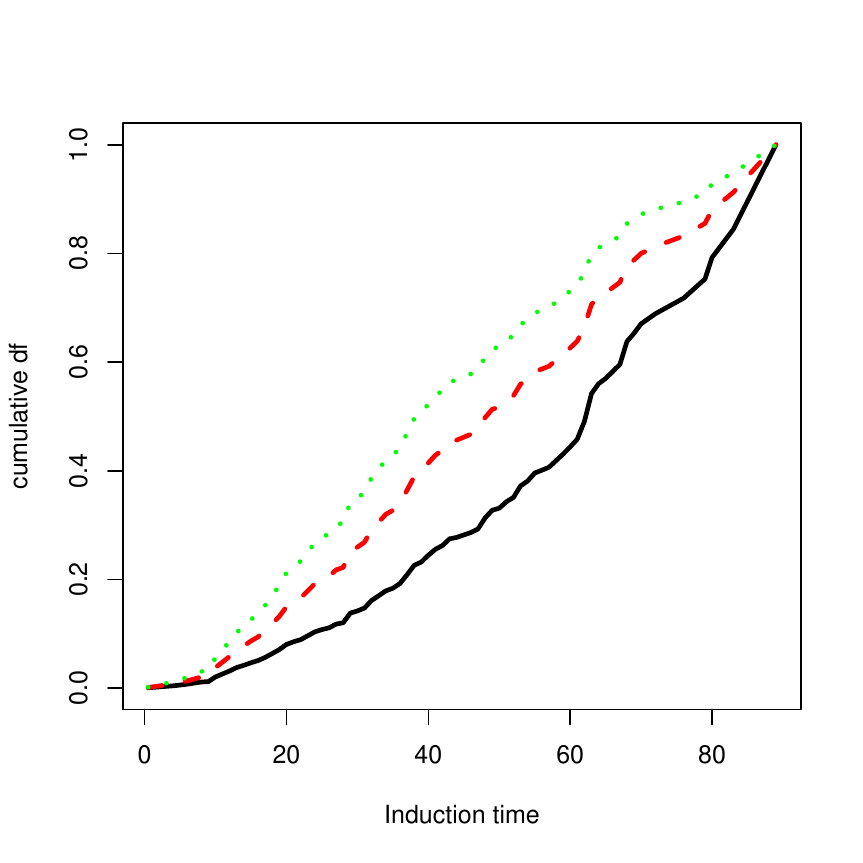}
\end{minipage} \hfill
\begin{minipage}[b]{0.5\linewidth}
\includegraphics[width=\textwidth]{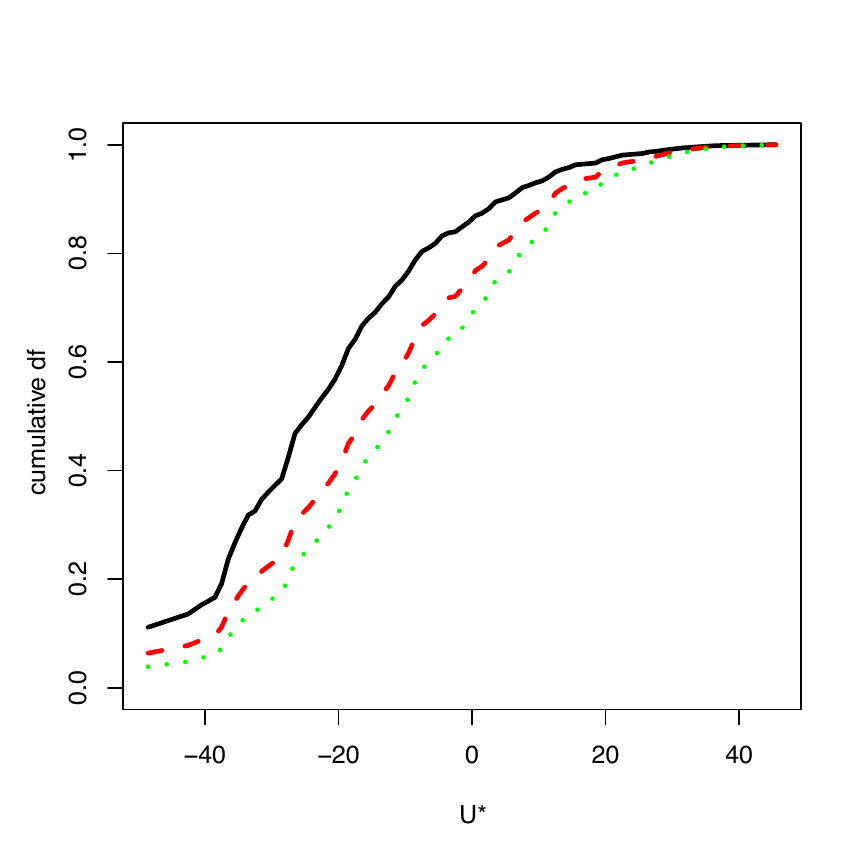}
\end{minipage} \hfill
\begin{minipage}[b]{0.5\linewidth}
\includegraphics[width=\textwidth]{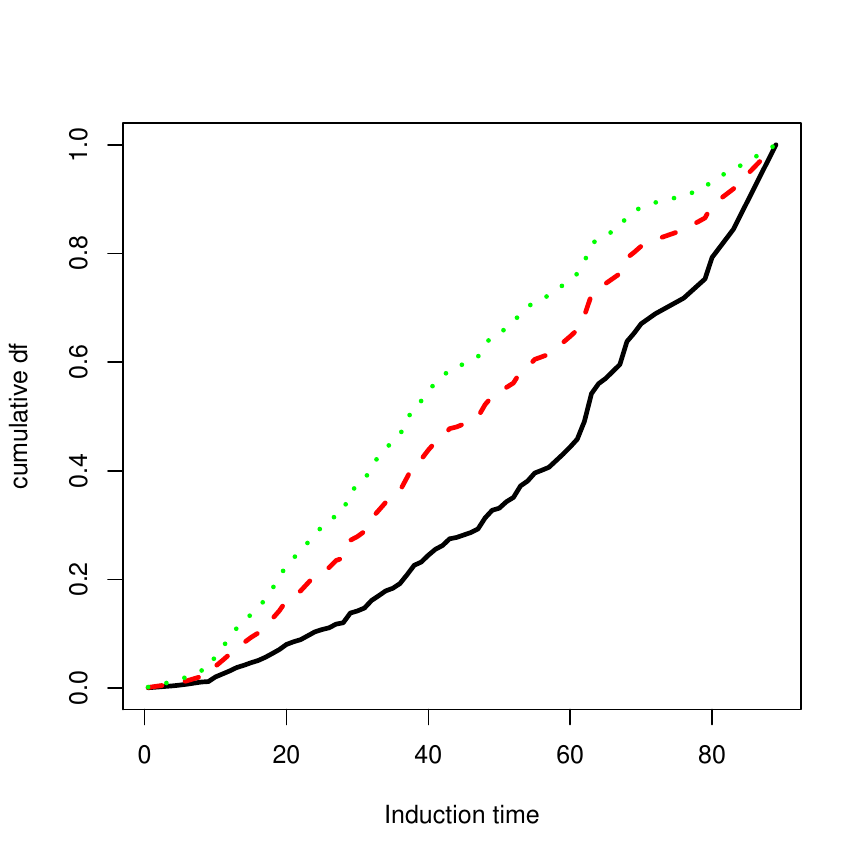}
\end{minipage} \hfill
\begin{minipage}[b]{0.5\linewidth}
\includegraphics[width=\textwidth]{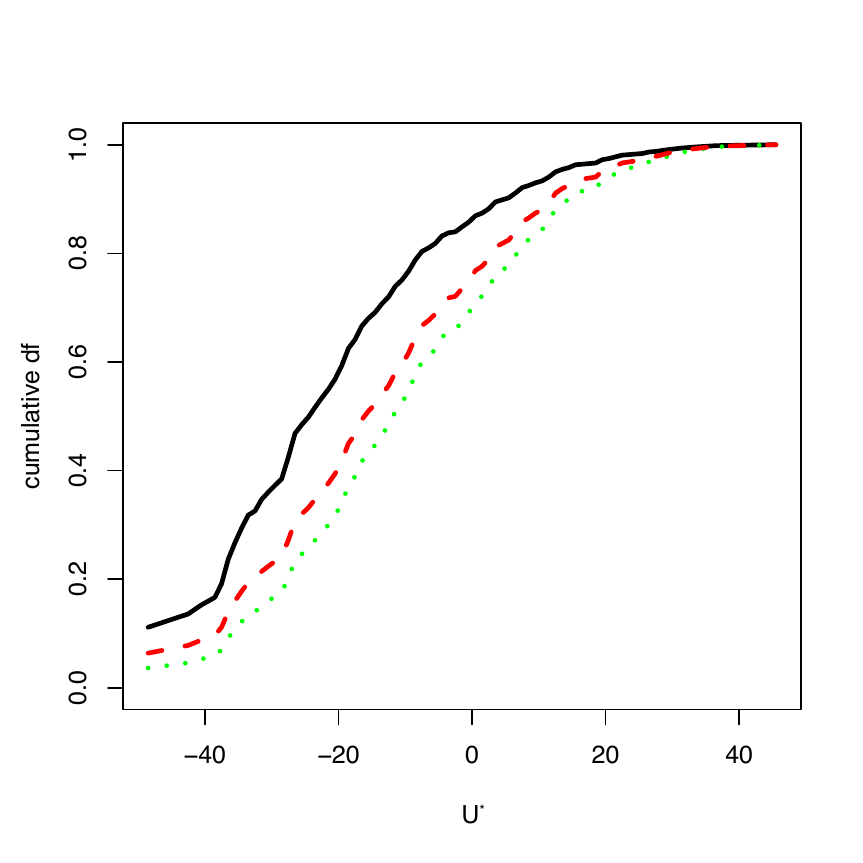}
\end{minipage} \hfill
\vspace{-0.2cm}
\vspace{-0.2cm}
\caption{ Cumulative distribution function for the incubation times (left) and the truncation time $U$ (right) using FGM copula (red dashed line) and Frank copula (green dotted line) and the NPMLE of Efron and Petrosian (black solid line). Simple algorithm (top) and Full algorithm (bottom). AIDS data.}
\label{Simple_model}
\end{figure}

In Figure \ref{Simple_model} (simple and full algorithms)
the cumulative df for the incubation times (left panels) and the truncation time $U^*$(right panels)
using the two copulas and the NPMLE under independence are jointly depicted. From this figure it is seen that the choice of the copula has some influence in the resulting estimator; however, the two copulas agree when shifting the Efron-Petrosian NPMLE of $F$ (left panel) and $G$ (right panel), although the correction entailed by the Frank copula is stronger. In this aspect, we note that the Frank copula function is able to take the full association between the incubation time and the truncation time into account. The FGM-copula  is restricted by its limited parameter space and therefore delivers a result between the result of the Frank copula and the independence setting. We also performed the analysis for the Clayton copula (results not shown). The full algorithm based on the Clayton copula  failed to provide a likely value for $\theta$. A possible explanation for this is that the full algorithm is unable to get away from the initial values of $(\theta,F,G)$ (the ones corresponding to the independent setting) when using this particular copula. The second and third order derivatives of the  Clayton copula are unbounded when $u_1$, $u_2$ approach zero, so, for small values of $F$ and $G$, the different weights $W$ containing these second order derivatives get large and dominate the likelihood function and also the optimum. This suggests a numerical issue probably related to the instability of the third-order derivatives of the Clayton copula around zero. For the AIDS data, the number of iterations needed to achieve the optimal value for the Clayton copula was much smaller than for the other copula functions. This  indicates that possibly a local optimum was reached instead of the global optimum. These limitations of the Clayton copula did not occur with the simple algorithm, which provided results consistent with those of the FGM and Frank copulas.

\begin{table}[!ht]
\caption {Number of iterations, estimated $\theta$, the correspondent Kendall's $\tau$, the standard error and the confidence interval for $\widehat \theta$,  using both algorithms, simple (top) and full (bottom). AIDS data.} \label{Table5}
\centering
\begin{tabular}{c r c c c c}
\hline
Copula& n. iter&$\widehat \theta$&SEboot&Interval&Kendall's $\tau$ \\
\cline{1-6}
FGM&55&0.982&0.3273&(0.3404;1.6235)&0.22\\
%Clayton&114&0.487&0.0785&(0.3330;0.6408)&0.20\\
Frank&179&3.350&0.7758&(1.8294;4.8706)&0.38\\
\hline
\hline
FGM&131&1.000&0.2425&(0.5246;1.4754)&0.22\\
%Clayton&25&0.07&0.0584&(-0.0445;0.1845)&0.03\\
Frank&186&3.460&0.6452&(2.1954;4.7245)&0.38\\
\hline
\end{tabular}
\end{table}
\vspace{-0.3cm}

%\begin{figure}[!h]
%\begin{minipage}[b]{0.45\linewidth}
%\includegraphics[width=\textwidth]{XvsCUMF_AIDS_Full_FGMvsClaytonvsFrank}
%\end{minipage} \hfill
%\begin{minipage}[b]{0.45\linewidth}
%\includegraphics[width=\textwidth]{UvsCUMK_AIDS_Full_FGMvsClaytonvsFrank}
%\end{minipage} \hfill
%\vspace{-0.2cm}
%\caption{ Cumulative distribution function for the incubation times (left) and the truncation time $U$ (right) using FGM copula (red dashed line), Clayton copula (blue dashed line), Frank %copula (green dashed line) and the NPMLE of Efron and Petrosian (black solid line). Full algorithm. AIDS data.}
%\label{Full_model}
%\end{figure}

\subsection{Immune Reconstitution Inflammatory Syndrome}
\label{Subsection42}

Data on Immune Reconstitution Inflammatory Syndrome (IRIS) were  collected from January 2008 to December 2017 in tertiary care teaching hospital at northern region of Portugal.  43 adult patients with HIV infection who first initiated antiretroviral therapy (TARV) and with CD4 cell counts $< 200/mmc$ or an AIDS-defining illness developed IRIS during that period. The time from TARV to IRIS diagnosis ($X^*$) is doubly truncated  by the time from beginning of the study (January 2008) and TARV initiation ($U^*$) and the time from TARV initiation and the end of the study, December 2017 ($V^*$). The difference between $U^*$ and its respective $V^*$ is always 10 years.\\

We have tested the quasi-independence between the time from TARV to IRIS diagnosis and the TARV initiation, based on the conditional Kendall's Tau of Martin and Betensky \cite{Betensky05}, and the quasi-independence assumption was widely accepted ($p =0.98$).
To confirm this finding we report in Table \ref{Table6} the value of $\widehat \theta$ (as well as the corresponding Kendall's Tau $\tau_\theta$) obtained from the two proposed algorithms (full  and simple),
for three copula families (FGM, Clayton and Frank). The number of iterations needed for the convergence of each algorithm  are included. Bootstrap standard errors and 95\% confidence intervals based on the bootstrap and the normal approximation are reported too, indicating no association between $U^*$ and $X^*$ in all cases. Therefore, there is no need to correct the Efron-Petrosian estimator for the IRIS data analysis. Indeed, the corrected estimator based on the FGM, Clayton and Frank copulas overlaps the Efron-Petrosian NPMLE in this case (results not shown).

\begin{table}[!ht]
\caption {Number of iterations, estimated $\theta$, the correspondent Kendall's $\tau$, the standard error and the confidence interval for $\widehat \theta$,  using both algorithms, simple (top) and full (bottom). IRIS data.} \label{Table6}
\centering
\begin{tabular}{c r c c c c}
\hline
Copula& n. iter&$\widehat \theta$&SEboot&Interval&Kendall's $\tau$ \\
\cline{1-6}
FGM&11&-0.059&0.716&(-1.461;1.342)&-0.006\\
Clayton&12&0.257&0.325&(-0.360;0.881)&0.110\\
Frank&11&-0.089&1.780&(-3.590;3.414)&-0.009\\
\hline
\hline
FGM&17&-0.007&0.502&(-1.051;0.917)&-0.001\\
Clayton&16&0.286&0.263&(-0.614;0.419)&0.030\\
Frank&14&-0.097&1.245&(-2.537;2.342)&-0.011\\
\hline
\end{tabular}
\end{table}
\vspace{-0.3cm}

\section{Conclusions and final discussion}
\label{sec:Section5}

In this paper we have introduced an extension of the Efron-Petrosian NPMLE for double truncation when
the variable of interest and the truncation variables may be dependent.  We assume that
$U^*$ depends on the variable of interest, and that the dependence structure of $(X^*,U^*)$ is given by a copula function, with arguments $\theta$, $F$ and $G$.
Two different algorithms to estimate the parameter  $\theta$ and the distributions $F$ and $G$ have been introduced, the full and the simple algorithms.\\

The performance of these two algorithms has been evaluated though simulations for several copula functions and marginal models. The estimators (of $F$) computed by the two algorithms converge to the same solution. While the simple algorithm provides MSE's slightly larger than those of the full algorithm, the full algorithm has revealed computationally heavier. The evaluations of the $RMSE$'s allows to concluded that the simple algorithm is the best option according to its good relative performance and computational speed. The systematic bias of the Efron-Petrosian NPMLE under dependence has been evaluated too, being more evident for a stronger dependence degree, as expected.\\

In order to estimate the standard error of both the marginal distributions and the copula parameter we have introduced a bootstrap procedure. In our simulation studies the bootstrap performed well, giving a more accurate estimation of the standard error of $\widehat \theta$ with an increasing sample size.\\

Two real data illustration have been provided.
 We have applied both algorithms to correct the Efron-Petrosian estimator of $F$ for the possible dependence between AIDS incubation time and the date of HIV infection, for different copula families. The copulas indicated a positive association between $U^*$ and $X^*$ when applying both algorithms. For the IRIS data, the proposed method based on copula function indicated no association between the time from TARV to IRIS diagnosis and the TARV initiation, and no correction of the Efron-Petrosian estimator was needed.\\
 The choice of the copula family is important since it has an impact in the final estimator. A possible approach for copula selection is an information criterion such as the AIC. For the AIDS data, were the correction for dependence is relevant, the minimum AIC was achieved by the Frank copula. This is in agreement with the discussion in Section \ref{subsection41}, in which the Frank copula was found to be more suitable to depict the dependence structure between the truncation and target variables. Of course, there exist many other copula families that can be used to model for dependence (\cite{Nelsen2006}). Formal goodness-of-fit tests for the copula model are missing however; this is an interesting topic for future research.

\section*{Appendix}
\label{Section6}

\subsection*{Appendix A.1: Likelihood calculations and score equations}

With the notations in Section \ref{sec:Section2}, the joint density of $(X^*,U^*)$ conditionally on $U^* \leq X^* \leq U^*+ \phi$ at point $(x,u)$ is given by
\begin{eqnarray*}   \frac{\mathcal{C}_{\theta}^{(1,1)}\left( F(x),K(u) \right)f(x) k(u)}{ \displaystyle \int_{u\leq x \leq u+ \phi} \displaystyle \int \mathcal{C}_{\theta}^{(1,1)}\left( F(x),K(u) \right)f(x)k(u)dxdu  },
\end{eqnarray*}

\noindent where $f(x)$ and $k(u)$ denote the densities corresponding to $F(x)$ and $K(u)$, respectively.
%This likelihood can be written as a product of the conditional likelihoods as:\\

%\begin{equation}
%L(\theta, f, k)= \displaystyle \prod_{i=1}^n \frac{W_{ii}f_i k_i}{\displaystyle \sum_{j=1}^n \displaystyle \sum_{m=1}^n W_{jm}f_j k_m I_{\left[ u_m \leq x_j \leq u_m+ \phi \right]}}
%\label{LikLH}
%\end{equation}

\noindent This justifies the likelihood (\ref{eq0}). In order to get the NPMLE for $\theta, f$ and $k$, we maximize the likelihood function (\ref{eq0}) , under the constraints $\displaystyle \sum_{i=1}^nf_i=1$ and $\displaystyle \sum_{i=1}^nk_i=1$. The loglikelihood is given by \\

\small{
\begin{equation}
log L(\theta, f, k)= \displaystyle \sum_{i=1}^n \left[log(f_i)+log (k_i)+log (W_{ii}^{(1,1)})-log \left( \displaystyle \sum_{j=1}^n \displaystyle \sum_{m=1}^nW_{jm}^{(1,1)}f_jk_mJ_{mj}\right) \right],
\end{equation}}

\noindent from which
\small{
\begin{eqnarray*}
&& \frac{\partial log L(\theta, f, k)}{\partial {f_m}}=\\
&=& \frac1{f_m}+ \displaystyle \sum_{i=1}^n \frac {W_{ii}^{(2,1)}}{W_{ii}^{(1,1)}}I_{[X_m \leq X_i ]}-\displaystyle \sum_{i=1}^n \frac{\displaystyle \sum_{j=1}^n \displaystyle \sum_{l=1}^n \left[ W_{jl}^{(2,1)}I_{[X_m \leq X_j]}f_jk_l J_{lj}+W_{jl}^{(1,1)}I_{[X_m = X_j] }k_l J_{lj}\right]}{\displaystyle \sum_{j=1}^n \displaystyle \sum_{l=1}^n W_{jl}^{(1,1)}f_jk_lJ_{lj}}\\
%&=&\frac1{f_m}- \displaystyle \sum_{i=1}^n \frac {\displaystyle \sum_{k=1}^n W_{mk}^{(1,1)}k_kJ_{km}}{\displaystyle \sum_{j=1}^n \displaystyle \sum_{k=1}^n W_{jk}^{(1,1)}f_jk_kJ_{kj}}+
%\displaystyle \sum_{i=1}^n \frac {W_{ii}^{(2,1)}}{W_{ii}^{(1,1)}}I_{[x_m \leq x_i ]}\\
%&-&\displaystyle \sum_{i=1}^n \frac {\displaystyle \sum_{j=1}^n \displaystyle \sum_{k=1}^n W_{jk}^{(2,1)}I_{[x_m \leq x_j ]}f_jk_kJ_{kj}}{\displaystyle \sum_{j=1}^n \displaystyle %\sum_{k=1}^n W_{jk}^{(1,1)}f_jk_kJ_{kj}}, m=1, \ldots, n\\
&&=  \frac1{f_m}+ \displaystyle \sum_{i=1}^n \frac { W_{ii}^{(2,1)}I_{[X_m\leq X_i]}}{ W_{ii}^{(1,1)}}-
\displaystyle \sum_{i=1}^n \frac {\displaystyle \sum_{j=1}^n \displaystyle \sum_{l=1}^n W_{jl}^{(2,1)}f_jk_lJ_{lj}I_{[X_m \leq X_j]}+\displaystyle \sum_{l=1}^n W_{ml}^{(1,1)}k_lJ_{lm} }{\displaystyle \sum_{j=1}^n \displaystyle \sum_{l=1}^n W_{jl}^{(1,1)}f_jk_lJ_{lj}},
\end{eqnarray*}}

\noindent and similarly

\begin{eqnarray*}
&& \frac{\partial log L(\theta, f, k)}{\partial {k_m}}=\\
&=&  \frac1{k_m}+ \displaystyle \sum_{i=1}^n \frac { W_{ii}^{(1,2)}I_{[U_m\leq U_i]}}{ W_{ii}^{(1,1)}}-
\displaystyle \sum_{i=1}^n \frac {\displaystyle \sum_{j=1}^n \displaystyle \sum_{l=1}^n W_{jl}^{(1,2)}f_jk_lJ_{lj}I_{[U_m \leq U_l]}+\displaystyle \sum_{j=1}^n W_{jm}^{(1,1)}f_jJ_{mj} }{\displaystyle \sum_{j=1}^n \displaystyle \sum_{l=1}^n W_{jl}^{(1,1)}f_jk_lJ_{lj}}.
\end{eqnarray*}

Solving the equation $\frac{\partial log L(\theta, f, k)}{\partial {f_m}}= 0$ we get\\

\begin{eqnarray*}
\frac 1{f_m}=n \frac{\displaystyle \sum_{j=1}^n \displaystyle \sum_{l=1}^n  { W_{jl}^{(2,1)}f_j k_l J_{lj}I_{[X_m \leq X_j]}}+\displaystyle \sum_{l=1}^n W_{ml}^{(1,1)}k_lJ_{lm} }{\displaystyle \sum_{j=1}^n \displaystyle \sum_{l=1}^n W_{jl}^{(1,1)}f_jk_lJ_{lj}}-\displaystyle \sum_{i=1}^n \frac {W_{ii}^{(2,1)}I_{[X_m \leq X_i]}}{ W_{ii}^{(1,1)}}
\end{eqnarray*}

\noindent from which

\small{
\begin{eqnarray*}
\hspace{-2cm}&&f_m=\\
\hspace{-1cm}&=& \frac {\displaystyle \sum_{j=1}^n \displaystyle \sum_{l=1}^n W_{jl}^{(1,1)}f_jk_lJ_{lj}}{n \displaystyle \sum_{j=1}^n \displaystyle \sum_{l=1}^n W_{jl}^{(2,1)}f_j k_l J_{lj}I_{[X_m \leq X_j]}+ n  \displaystyle \sum_{l=1}^n  W_{ml}^{(1,1)}k_lJ_{lm}-{\displaystyle \sum_{j=1}^n \displaystyle \sum_{l=1}^n W_{jl}^{(1,1)}f_jk_lJ_{lj}}\displaystyle \sum_{i=1}^n \frac { W_{ii}^{(2,1)}I_{[X_m\leq X_i]}}{ W_{ii}^{(1,1)}}}
\end{eqnarray*}}

\begin{eqnarray*}
\hspace{-1cm}&=& \frac \alpha {n A_m+ n  K_m^{W}-\alpha B_m}.
\end{eqnarray*}

\noindent Since $\displaystyle \sum_{m=1}^n f_m=1$, we get that
\begin{eqnarray*}
\alpha \displaystyle \sum_{m=1}^n \frac 1{n A_m+ n  K_m^{W}-\alpha B_m}=1
\end{eqnarray*}.

\noindent Then,

\begin{eqnarray*}
\alpha= \left[\displaystyle \sum_{m=1}^n \frac 1{n A_m+ n  K_m^{W}-\alpha B_m} \right]^{-1}.
\end{eqnarray*}

This proves the score equation (\ref{eq1full}).

To justify the score equation (\ref{eq2full}), from $\frac {\partial log L (\theta, f, k)}{\partial k_m}=0$,  we have similarly:\\

\begin{eqnarray*}
 \frac 1{k_m}= n\frac{\displaystyle \sum_{j=1}^n \displaystyle \sum_{l=1}^n  { W_{jl}^{(1,2)}f_j k_l J_{lj}I_{[U_m \leq U_k]}}+ { \displaystyle \sum_{j=1}^n W_{jm}^{(1,1)}f_j J_{mj}}}{\alpha}\\
&-&\displaystyle \sum_{i=1}^n \frac {W_{ii}^{(1,2)}I_{[U_m \leq U_i]}}{ W_{ii}^{(1,1)}}.
\end{eqnarray*}

From this equation and since $\displaystyle \sum_{m=1}^n k_m=1$, we get (\ref{eq2full}).

\begin{eqnarray*}
k_h&=& \left[\displaystyle \sum_{m=1}^n \frac 1{n \displaystyle \sum_{j=1}^n \displaystyle \sum_{l=1}^n W_{jl}^{(2,1)}f_j k_l J_{lj}I_{[X_m \leq X_i]}+ n  \displaystyle \sum_{l=1}^n  W_{ml}^{(1,1)}k_lJ_{lm}-\alpha\displaystyle \sum_{i=1}^n \frac { W_{ii}^{(2,1)}I_{[X_m\leq X_i]}}{ W_{ii}^{(1,1)}}} \right]^{-1} \\
&& \times \frac 1 n{ \displaystyle \sum_{j=1}^n \displaystyle \sum_{l=1}^n W_{jl}^{(1,2)}f_j k_l J_{lj}I_{[U_h \leq U_j]}+  n\displaystyle \sum_{j=1}^n  W_{jh}^{(1,1)}f_jJ_{hj}- \alpha \displaystyle \sum_{i=1}^n \frac { W_{ii}^{(1,2)}I_{[U_h\leq U_i]}}{ W_{ii}^{(1,1)}}}
\end{eqnarray*}\\

\subsection*{Appendix A.2: Identifiability of the copula model}

Assume $P(X^*\leq x, U^* \leq u)=\mathcal{C}(F(x),K(u))$, $(x,u) \in [a_F,b_F] \times [a_G,b_G]$, where $\mathcal{C}$ is a copula function and $[a_F,b_F]$ and $[a_G,b_F]$ denote the supports of $X^*$ and $U^*$. For identifiability of $F$ and $K$ assume $a_F-\phi \leq a_G\leq a_F$ and $b_G\leq b_F\leq b_F+\phi$ (\cite{Woodroofe85}). The joint density of $(X^*,U^*)$ is given by $f(x,u)=\mathcal{C}^{(1,1)}(F(x),K(u))f(x)k(u)$ where $\mathcal{C}^{(1,1)}(w_1,w_2)=\partial ^2\mathcal{C}(w_1,w_2)/\partial w_1 \partial w_2$ is the copula density and $f(x)$ and $k(u)$ denote the densities of $F$ and $K$. On the other hand, due to the interval sampling, the joint density of the observable variables is $\tilde f(x,u)=\mathcal{C}^{(1,1)}(F(x),K(u))f(x)k(u)/P(U^*\leq X^*\leq U^* + \phi)$, $u\leq x\leq u+\phi$. Note that when one changes $f(x,u)$ for $x<u$ (lower right corner) this does not affect $\tilde f$ but leads to a different copula function and attached marginals in the copula model. Hence, as in Ding \cite{Ding2012}, we introduce an identifiability condition for $\mathcal{C}$, where the notation $h^{(1)}$ stands for the first derivative of the function $h$.\\

\textbf{Definition.} A copula family $\mathcal{F}=\{\mathcal{C}(w_1,w_2), 0\leq w_1,w_2\leq 1\}$ is called double truncated identifiable if for $\mathcal{C},\mathcal{C}^* \in \mathcal{F}$ and differentiable non-decreasing functions $w^*_1(w_1),w_2^*(w_2)$,

\begin{equation}
\mathcal{C}^{(1,1)}(w_1,w_2)=b \times \mathcal{C}^{*(1,1)}(w_1^*(w_1),w_2^*(w_2))w_1^{*(1)}(w_1)w_2^{*(1)}(w_2)
\label{eq:identif}
\end{equation}

\noindent for a constant $b$ on an open set in $\{(w_1,w_2),0\leq w_1\leq w_2,0\leq a_1(w_1)\leq w_2\leq a_2(w_1)\leq 1\}$, with $a_1(w_1),a_2(w_1)$ non-decreasing functions on $[0,1]$, implies that $\mathcal{C}=\mathcal{C}^*$, $b=1$, $w_1^*(w_1)=w_1$ and $w_2^*(w_2)=w_2$.\\

In order to prove that the Definition above indeed allows for an identifiable copula model with interval sampling, take two sets of functions $\mathcal{C}$, $F$, $f$, $K$, $k$ and $\mathcal{C}^*$, $F^*$, $f^*$, $K^*$, $k^*$ leading to the same $\tilde f(x,u)$. Then, on $u\leq x\leq u+\phi$,

$$\mathcal{C}^{(1,1)}(F(x),K(u))=\frac{c}{c^*} \times \mathcal{C}^{*(1,1)}(F^*(x),K^*(u))f^*(x)k^*(u)/f(x)k(u),$$

\noindent where (in an obvious notation) $c=P(U^*\leq X^*\leq U^*+\phi)$ and $c^*=P^*(U^*\leq X^*\leq U^*+\phi)$. Introduce $b=c/c^*$, $w_1^*(w_1)=K^*(K^{-1}(w_1))$, $w_2^*(w_2)=F^*(F^{-1}(w_2))$. Then, (\ref{eq:identif}) holds with $a_1(w_1)=F(G^{-1}(w_1))$ and $a_2(w_1)=F(K^{-1}(w_1)+\phi)$. The Definition above implies $\mathcal{C}=\mathcal{C}^*$, $F=F^*$, $K=K^*$ and the copula model is identifiable, as announced.

%\begin{eqnarray*}
%\hat K(x)&=& \displaystyle \sum_{m=1}^n \left[\displaystyle \sum_{m=1}^n \frac 1{n \displaystyle \sum_{j=1}^n \displaystyle \sum_{k=1}^n W_{jk}^{(2,1)}f_j k_k J_{kj}I_{[x_m \leq x_i]}+ n  %\displaystyle \sum_{k=1}^n  W_{mk}^{(1,1)}k_kJ_{km}-\alpha\displaystyle \sum_{i=1}^n \frac { W_{ii}^{(2,1)}I_{[x_m\leq x_i]}}{ W_{ii}^{(1,1)}}} \right]^{-1} \\
%&& \times \frac {I_{[u_m \leq u_i]}} {n \displaystyle \sum_{j=1}^n \displaystyle \sum_{k=1}^n W_{jk}^{(2,1)}f_j k_k J_{kj}I_{[x_k \leq x_j]}+  \displaystyle \sum_{k=1}^n  %W_{mk}^{(1,1)}k_kJ_{km}- \alpha \displaystyle \sum_{i=1}^n \frac { W_{ii}^{(2,1)}I_{[x_m\leq x_i]}}{ W_{ii}^{(1,1)}}}
%\end{eqnarray*}

\subsection*{Appendix A.3: Simulation results for small sample sizes}

\begin{table}[ht]
\caption {MSE's of the proposed estimators $\widehat F$ and $\widehat G$, in each decile and each copula's family, for the full algorithm and $n=50$ and $n=100$.}
\centering
\begin{tabular}{ccrcccccccccc}
  \hline
Copula& $n$& $\theta$ & & 0.1 & 0.2 & 0.3 & 0.4 & 0.5 & 0.6 & 0.7 & 0.8 & 0.9 \\
  \hline
& & -1&  & 0.00281 & 0.00551 & 0.00656 & 0.00712 & 0.00669 & 0.00590 & 0.00499 & 0.00357 & 0.00206 \\
&&-0.5&$\widehat F$ &  0.00331 & 0.00588 & 0.00673 & 0.00703 & 0.00645 & 0.00559 & 0.00469 & 0.00335 & 0.00197 \\
&&1&& 0.00233 & 0.00424 & 0.00511 & 0.00567 & 0.00551 & 0.00478 & 0.00417 & 0.00306 & 0.00179 \\
\cline{3-13}
 &50&-1& &  0.00202 & 0.00354 & 0.00454 & 0.00549 & 0.00615 & 0.00657 & 0.00622 & 0.00548 & 0.00347 \\
&& -0.5&$\widehat G$ & 0.00204 & 0.00331 & 0.00427 & 0.00534 & 0.00630 & 0.00692 & 0.00682 & 0.00604 & 0.00393 \\
&&1&&  0.00184 & 0.00311 & 0.00393 & 0.00454 & 0.00483 & 0.00515 & 0.00509 & 0.00463 & 0.00286 \\
\cline{2-13}
FGM&&-1& & 0.00172 & 0.00297 & 0.00354 & 0.00352 & 0.00323 & 0.00292 & 0.00246 & 0.00180 & 0.00099 \\
&&-0.5&$\widehat F$ & 0.00188 & 0.00312 & 0.00366 & 0.00354 & 0.00324 & 0.00292 & 0.00245 & 0.00173 & 0.00095 \\
&&1&& 0.00116 & 0.00193 & 0.00231 & 0.00246 & 0.00243 & 0.00229 & 0.00204 & 0.00152 & 0.00084 \\
\cline{3-13}
&100&-1&& 0.00099 & 0.00184 & 0.00224 & 0.00269 & 0.00296 & 0.00322 & 0.00338 & 0.00294 & 0.00183 \\
&&-0.5&$\widehat G$ & 0.00096 & 0.00165 & 0.00219 & 0.00264 & 0.00291 & 0.00328 & 0.00354 & 0.00308 & 0.00194 \\
&&1&& 0.00082 & 0.00161 & 0.00195 & 0.00224 & 0.00239 & 0.0026 & 0.00243 & 0.00202 & 0.00119 \\
\hline
   & & -2.1&   & 0.00222 & 0.00454 & 0.00531 & 0.00584 & 0.00535 & 0.00478 & 0.00499 & 0.00327 & 0.00189 \\
&&-1& & 0.00281 & 0.00551 & 0.00656 & 0.00712 & 0.00669 & 0.00590 & 0.00499 & 0.00357 & 0.00206 \\
&&1.86&$\widehat F$ & 0.00233 & 0.00424 & 0.00511 & 0.00567 & 0.00551 & 0.00478 & 0.00417 & 0.00306 & 0.00179 \\
&&5.74&& 0.00331 & 0.00588 & 0.00673 & 0.00703 & 0.00645 & 0.00559 & 0.00469 & 0.00335 & 0.00197 \\
&&20.9&& 0.00198 & 0.00379 & 0.00418 & 0.00455 & 0.00517 & 0.00474 & 0.00378 & 0.00327 & 0.00174 \\
\cline{3-13}
 &50&-2.1&   & 0.00212 & 0.00314 & 0.00424 & 0.00449 & 0.00515 & 0.00557 & 0.00522 & 0.00448 & 0.00247 \\
&& -1&  & 0.00202 & 0.00354 & 0.00454 & 0.00549 & 0.00615 & 0.00657 & 0.00622 & 0.00548 & 0.00347 \\
&&1.86& $\widehat G$& 0.00184 & 0.00311 & 0.00393 & 0.00454 & 0.00483 & 0.00515 & 0.00509 & 0.00463 & 0.00286 \\
&&5.74& & 0.00204 & 0.00331 & 0.00427 & 0.00534 & 0.0063 & 0.00692 & 0.00682 & 0.00604 & 0.00393 \\
&&20.9& & 0.00180 & 0.00312 & 0.00410 & 0.00492 & 0.00510 & 0.00502 & 0.00453 & 0.00316 & 0.00201 \\
\cline{2-13}
Frank&&-2.1&  & 0.00112 & 0.00161 & 0.00208 & 0.00219 & 0.00235 & 0.00262 & 0.00198 & 0.00239 & 0.00098 \\
&&-1&  & 0.00172 & 0.00297 & 0.00354 & 0.00352 & 0.00323 & 0.00292 & 0.00246 & 0.0018 & 0.00099 \\
&&1.86&$\widehat F$ & 0.00106 & 0.00194 & 0.00228 & 0.00237 & 0.00244 & 0.00235 & 0.00200 & 0.00156 & 0.00088 \\
&&5.74&& 0.00118 & 0.00199 & 0.00242 & 0.00261 & 0.00275 & 0.00231 & 0.00196 & 0.00148 & 0.00087 \\
&&20.9&& 0.00920 & 0.00142 & 0.00185 & 0.00231 & 0.00265 & 0.00231 & 0.00206 & 0.00188 & 0.00085 \\
\cline{3-13}
&100&-2.1& & 0.00109 & 0.00171 & 0.00226 & 0.00251 & 0.00263 & 0.00203 & 0.00233 & 0.00194 & 0.00135 \\
&&-1& & 0.00099 & 0.00184 & 0.00224 & 0.00269 & 0.00296 & 0.00322 & 0.00338 & 0.00294 & 0.00183 \\
&&1.86&$\widehat G$& 0.00105 & 0.00167 & 0.00203 & 0.00226 & 0.00232 & 0.00245 & 0.00242 & 0.00198 & 0.00120 \\
&&5.74&& 0.00089 & 0.00166 & 0.00211 & 0.00241 & 0.00258 & 0.00266 & 0.00253 & 0.00194 & 0.00130 \\
&&20.9&& 0.00092 & 0.00146 & 0.00201 & 0.00221 & 0.00238 & 0.00246 & 0.00213 & 0.00154 & 0.00092 \\
   \hline
      & & 0.5&   & 0.00202 & 0.00354 & 0.00431 & 0.00484 & 0.00530 & 0.00476 & 0.00399 & 0.00297 & 0.00169 \\
&&2&$\widehat F$ & 0.00179 & 0.00341 & 0.00427 & 0.00478 & 0.00498 & 0.00475 & 0.00417 & 0.00300 & 0.00177 \\
&&18&& 0.00198 & 0.00389 & 0.00488 & 0.00545 & 0.00547 & 0.00484 & 0.00378 & 0.00293 & 0.00164 \\
\cline{3-13}
 &50&0.5&   & 0.00202 & 0.00354 & 0.00454 & 0.00549 & 0.00615 & 0.00657 & 0.00622 & 0.00548 & 0.00347 \\
&& 2&$\widehat G$  & 0.00167 & 0.00295 & 0.00373 & 0.00439 & 0.00472 & 0.00486 & 0.00459 & 0.00337 & 0.00217 \\
&&18& & 0.00183 & 0.00308 & 0.00410 & 0.00472 & 0.00505 & 0.0053 & 0.00514 & 0.00423 & 0.00244 \\
\cline{2-13}
Clayton&&0.5&  & 0.00110 & 0.00171 & 0.00228 & 0.00239 & 0.00245 & 0.00232 & 0.00191 & 0.00139 & 0.00085 \\
&&2&$\widehat F$   & 0.00107 & 0.00170 & 0.00229 & 0.00242 & 0.00245 & 0.00235 & 0.00195 & 0.00144 & 0.00087 \\
&&18&& 0.00118 & 0.00199 & 0.00242 & 0.00261 & 0.00275 & 0.00231 & 0.00196 & 0.00148 & 0.00087 \\
\cline{3-13}
&100&0.5& & 0.00089 & 0.00161 & 0.00206 & 0.00221 & 0.00243 & 0.00243 & 0.00221 & 0.00174 & 0.00105 \\
&&2&$\widehat G$ & 0.00077 & 0.00132 & 0.00181 & 0.00223 & 0.00286 & 0.00335 & 0.00341 & 0.00279 & 0.00164 \\
&&18&& 0.00089 & 0.00166 & 0.00211 & 0.00241 & 0.00258 & 0.00266 & 0.00253 & 0.00194 & 0.00120 \\
   \hline

\end{tabular}
\end{table}

\begin{table}[ht]
\caption {MSE's of the proposed estimators $\widehat F$ and $\widehat G$, in each decile and each copula's family, for the simple algorithm for different $\theta$'s and $n=50$ and $n=100$.}
\centering
\begin{tabular}{ccrcccccccccc}
  \hline
Copula& $n$& $\theta$ & & 0.1 & 0.2 & 0.3 & 0.4 & 0.5 & 0.6 & 0.7 & 0.8 & 0.9 \\
  \hline
& & -1&   & 0.00303 & 0.00585 & 0.00687 & 0.00744 & 0.00693 & 0.00604 & 0.00503 & 0.00358 & 0.00207 \\
&&-0.5&$\widehat F$  & 0.00339 & 0.00606 & 0.00694 & 0.00728 & 0.00665 & 0.00571 & 0.00473 & 0.00335 & 0.00197 \\
&&1& & 0.00233 & 0.00424 & 0.00507 & 0.00562 & 0.00542 & 0.00476 & 0.00418 & 0.00308 & 0.00183 \\
\cline{3-13}
 &50&-1& & 0.00190 & 0.00354 & 0.00471 & 0.00588 & 0.00671 & 0.00719 & 0.00674 & 0.00582 & 0.00359 \\
&& -0.5&$\widehat G$  & 0.00197 & 0.00335 & 0.00441 & 0.00561 & 0.00662 & 0.00725 & 0.00711 & 0.00626 & 0.00398 \\
&&1&& 0.00178 & 0.00311 & 0.00405 & 0.00477 & 0.00505 & 0.00534 & 0.00515 & 0.00458 & 0.00274 \\
\cline{2-13}
FGM&&-1& & 0.00181 & 0.00314 & 0.00376 & 0.00374 & 0.00342 & 0.00307 & 0.00257 & 0.00187 & 0.00101 \\
&&-0.5&$\widehat F$ & 0.00190 & 0.00313 & 0.00369 & 0.00357 & 0.00326 & 0.00294 & 0.00247 & 0.00175 & 0.00095 \\
&&1& & 0.00156 & 0.00265 & 0.00316 & 0.00317 & 0.00304 & 0.00273 & 0.00233 & 0.00167 & 0.00091 \\
\cline{3-13}
&100&-1&& 0.00092 & 0.00184 & 0.00236 & 0.00292 & 0.00330 & 0.00360 & 0.00375 & 0.00316 & 0.00182 \\
&&-0.5&$\widehat G$ & 0.00094 & 0.00168 & 0.00227 & 0.00277 & 0.00303 & 0.00339 & 0.0036 & 0.00307 & 0.00186 \\
&&1& & 0.00090 & 0.00155 & 0.00220 & 0.00253 & 0.00285 & 0.00306 & 0.00315 & 0.00262 & 0.00156 \\
\hline
   & & -2.1&   & 0.00958 & 0.01420 & 0.01487 & 0.01317 & 0.01081 & 0.00818 & 0.00596 & 0.00393 & 0.00196 \\
&&-1&  & 0.00748 & 0.01085 & 0.01116 & 0.01028 & 0.00877 & 0.00685 & 0.00521 & 0.00364 & 0.00189 \\
&&1.86&$\widehat F$ & 0.00240 & 0.00394 & 0.00461 & 0.00506 & 0.00473 & 0.00449 & 0.00397 & 0.00312 & 0.00173 \\
&&5.74& & 0.00187 & 0.00333 & 0.00408 & 0.00442 & 0.00467 & 0.00459 & 0.00383 & 0.003 & 0.00187 \\
&&20.9&& 0.00190 & 0.00316 & 0.00410 & 0.00470 & 0.00485 & 0.00458 & 0.00404 & 0.00329 & 0.00182 \\
\cline{3-13}
 &50&-2.1&   & 0.00184 & 0.00383 & 0.00578 & 0.00840 & 0.01106 & 0.01355 & 0.01576 & 0.01607 & 0.01277 \\
&& -1&  & 0.00172 & 0.00356 & 0.00514 & 0.00708 & 0.00888 & 0.01047 & 0.01164 & 0.01153 & 0.00884 \\
&&1.86& $\widehat G$ & 0.00168 & 0.00312 & 0.00395 & 0.00491 & 0.00526 & 0.00499 & 0.00475 & 0.00386 & 0.00239 \\
&&5.74&  & 0.00171 & 0.00316 & 0.00401 & 0.00485 & 0.00510 & 0.00476 & 0.00433 & 0.00331 & 0.00198 \\
&&20.9& & 0.00172 & 0.00313 & 0.00398 & 0.00484 & 0.00508 & 0.00465 & 0.00420 & 0.00315 & 0.00199 \\
\cline{2-13}
Frank&&-2.1&  & 0.00498 & 0.00743 & 0.00748 & 0.00651 & 0.00533 & 0.00415 & 0.00298 & 0.00195 & 0.00095 \\
&&-1&  & 0.00305 & 0.00460 & 0.00492 & 0.00452 & 0.00379 & 0.00312 & 0.00244 & 0.00166 & 0.00085 \\
&&1.86&$\widehat F$  & 0.00104 & 0.00193 & 0.00230 & 0.00241 & 0.00247 & 0.00238 & 0.00202 & 0.00157 & 0.00088 \\
&&5.74& & 0.00086 & 0.00152 & 0.00203 & 0.00222 & 0.00236 & 0.00236 & 0.00204 & 0.00159 & 0.00094 \\
&&20.9& & 0.00093 & 0.00155 & 0.00196 & 0.00224 & 0.00240 & 0.00240 & 0.00209 & 0.00159 & 0.00086 \\
\cline{3-13}
&100&-2.1& & 0.00100 & 0.00194 & 0.00301 & 0.00420 & 0.00555 & 0.00698 & 0.00802 & 0.00786 & 0.00572 \\
&&-1& & 0.00092 & 0.00171 & 0.00249 & 0.00331 & 0.00405 & 0.00490 & 0.00538 & 0.00495 & 0.00322 \\
&&1.86&$\widehat G$ & 0.00090 & 0.00155 & 0.00203 & 0.00240 & 0.00252 & 0.00263 & 0.00251 & 0.00196 & 0.00111 \\
&&5.74&& 0.00094 & 0.00155 & 0.00204 & 0.00232 & 0.00233 & 0.00237 & 0.00217 & 0.00163 & 0.00093 \\
&&20.9& & 0.00096 & 0.00156 & 0.00202 & 0.00231 & 0.00233 & 0.00234 & 0.00214 & 0.00161 & 0.00091 \\
   \hline
         & & 0.5&    & 0.00164 & 0.00329 & 0.00422 & 0.00464 & 0.00502 & 0.0045 & 0.00402 & 0.00310 & 0.00173 \\
&&2&$\widehat F$  & 0.00202 & 0.00354 & 0.00431 & 0.00484 & 0.00530 & 0.00476 & 0.00399 & 0.00297 & 0.00169 \\
&&18& & 0.00214 & 0.00408 & 0.00474 & 0.00515 & 0.00525 & 0.00470 & 0.00385 & 0.00294 & 0.00167 \\
\cline{3-13}
 &50&0.5&    & 0.00167 & 0.00308 & 0.00430 & 0.00465 & 0.00485 & 0.00461 & 0.00413 & 0.0033 & 0.00182 \\
&& 2&$\widehat G$  & 0.00182 & 0.00312 & 0.00389 & 0.00449 & 0.00474 & 0.00496 & 0.00483 & 0.00367 & 0.00223 \\
&&18&  & 0.00178 & 0.00318 & 0.00421 & 0.00479 & 0.00500 & 0.00540 & 0.00515 & 0.00427 & 0.00245 \\
\cline{2-13}
Clayton&&0.5&  & 0.00088 & 0.00159 & 0.00202 & 0.00246 & 0.00245 & 0.00231 & 0.00197 & 0.00147 & 0.00087 \\
&&2&$\widehat F$   & 0.00110 & 0.00171 & 0.00228 & 0.00239 & 0.00245 & 0.00232 & 0.00191 & 0.00139 & 0.00085 \\
&&18&& 0.00125 & 0.00205 & 0.00246 & 0.00256 & 0.00266 & 0.00240 & 0.00193 & 0.00145 & 0.00088 \\
\cline{3-13}
&100&0.5& &0.00090 & 0.00156 & 0.00203 & 0.00232 & 0.00243 & 0.00241 & 0.00212 & 0.00163 & 0.00094 \\
&&2&$\widehat G$ & 0.00089 & 0.00161 & 0.00206 & 0.00221 & 0.00243 & 0.00243 & 0.00221 & 0.00174 & 0.00105 \\
&&18&  & 0.00096 & 0.00163 & 0.00216 & 0.00240 & 0.00262 & 0.00263 & 0.00241 & 0.00195 & 0.00115 \\
   \hline
\end{tabular}
\end{table}

\begin{table}[ht]
\caption {The bias and the standard deviation of the estimator $\widehat \theta$ obtained from the simple algorithm along the 1,000 trials, for each copula function, different $\theta$'s and sample sizes $n=50, 100$.}
\begin{center}
\begin{tabular}{cccrc}
  \hline
Copula& n& $\theta$& Bias($\widehat \theta$)& sd($\widehat \theta$)\\
 \hline
& & -1& 0.1856&0.3028\\
&50&-0.5&-0.0366&0.4633\\
&&1&-0.2209&0.3481\\
\cline{2-5}
FGM& & -1& 0.1321&0.2041\\
&100 & -0.5& 0.0331&0.3521\\
& & 1& -0.1225&0.1967\\
\hline
\hline
& & -2.1& -0.1995&1.5008\\
&&-1&-0.1266&1.4246\\
&50&1.86&0.0859&1.1165\\
&&5.74&0.1375&1.2837\\
&&20.9&-0.8447&3.0419\\
\cline{2-5}
Frank & & -2.1& -0.1529&1.1036\\
& & -1& -0.0743&0.9541\\
& 100& 1.86&0.0532&0.7854\\
& & 5.74&0.1098&0.9002\\
& & 20.9&-0.3990&2.1974\\
\hline
\hline
& & 0.5& 1.6045&0.6073\\
&50&2&-1.3871&0.3235\\
&&18&-17.8531&0.1901\\
\cline{2-5}
Clayton& & 0.5& 1.5517&0.4047\\
&100 & 2& -1.2312&0.2269\\
& & 18& -16.8930&0.1246\\
\hline
\hline
\end{tabular}
\label{Thetasmall}\end{center}
\end{table}

\begin{table}[ht]
\caption {Bias of the NMPLE proposed by Shen, in each decile, and each functions $F$ and $G$ , for FGM, Frank and Clayton copulas, with sample size
$n=50$ and $n=100$ and different $\tau$'s.}
\centering
\small{
\begin{tabular}{ccrrrrrrrrrrr}
  \hline
Copula& $n$& $\tau$ & & 0.1 & 0.2 & 0.3 & 0.4 & 0.5 & 0.6 & 0.7 & 0.8 & 0.9 \\
  \hline
& & -0.22&   & 0.02534 & 0.03838 & 0.04180 & 0.03774 & 0.03067 & 0.02367 & 0.01664 & 0.00960 & 0.00257 \\
&&-0.11&$ F$   & 0.01843 & 0.02642 & 0.02807 & 0.02522 & 0.02063 & 0.01604 & 0.01149 & 0.00694 & 0.00232 \\
&&0&& -0.00968 & -0.01547 & -0.01722 & -0.01555 & -0.01270 & -0.00990 & -0.00702 & -0.00412 & -0.00109 \\
&&0.22&  & -0.02335 & -0.03512 & -0.03803 & -0.03431 & -0.02813 & -0.02203 & -0.01576 & -0.00953 & -0.00324 \\
\cline{3-13}
 &50&-0.22& & -0.00260 & -0.00962 & -0.01668 & -0.02390 & -0.03108 & -0.03813 & -0.04222 & -0.03887 & -0.02559 \\
&& -0.11&$ G$  & -0.00249 & -0.00699 & -0.01177 & -0.01617 & -0.02072 & -0.02533 & -0.02838 & -0.02700 & -0.01881 \\
&&0& & 0.00114 & 0.00410 & 0.00694 & 0.0099 & 0.01270 & 0.01568 & 0.01728 & 0.01565 & 0.00984 \\
&&0.22& & 0.00334 & 0.00963 & 0.01578 & 0.02186 & 0.02810 & 0.03442 & 0.03813 & 0.03528 & 0.02337 \\
\cline{2-13}
FGM&&-0.22& & 0.02534 & 0.03838 & 0.04180 & 0.03774 & 0.03067 & 0.02367 & 0.01664 & 0.00960 & 0.00257 \\
&&-0.11&$F$ & 0.01667 & 0.02437 & 0.02613 & 0.02357 & 0.01924 & 0.01497 & 0.01069 & 0.00644 & 0.00212 \\
&&0& & 0.00133 & 0.00140 & 0.00124 & 0.00105 & 0.00086 & 0.00069 & 0.00054 & 0.00046 & 0.00031 \\
&&0.22&& -0.02229 & -0.03417 & -0.03564 & -0.03223 & -0.02653 & -0.02024 & -0.01373 & -0.00757 & -0.00205 \\
\cline{3-13}
&100&-22& & -0.00260 & -0.00962 & -0.01668 & -0.02390 & -0.03108 & -0.03813 & -0.04222 & -0.03887 & -0.02559 \\
&&-0.11&$ G$ & -0.00215 & -0.00645 & -0.01086 & -0.01515 & -0.01953 & -0.02386 & -0.0264 & -0.02455 & -0.01695 \\
&&0&  & -0.00031 & -0.00052 & -0.00067 & -0.00091 & -0.00105 & -0.00116 & -0.00128 & -0.00139 & -0.00136 \\
&&0.22&& 0.00253 & 0.00696 & 0.01303 & 0.01850 & 0.02383 & 0.03212 & 0.03586 & 0.03236 & 0.02103 \\
\hline
   & & -0.22&   & 0.05300 & 0.06978 & 0.07123 & 0.06258 & 0.05121 & 0.03970 & 0.02819 & 0.01672 & 0.00472 \\
&&-0.11&   & 0.03055 & 0.03920 & 0.03960 & 0.03492 & 0.02860 & 0.02249 & 0.01624 & 0.00989 & 0.00331 \\
&&0&$ F$ & 0.01789 & 0.02604 & 0.02782 & 0.02508 & 0.02059 & 0.01604 & 0.01151 & 0.00699 & 0.00228 \\
&&0.22& & -0.02256 & -0.03397 & -0.03754 & -0.03456 & -0.02857 & -0.02255 & -0.01669 & -0.01049 & -0.00406 \\
&&0.5&& -0.04592 & -0.07100 & -0.07961 & -0.07414 & -0.06255 & -0.05064 & -0.03864 & -0.02558 & -0.01172 \\
&&0.9&& -0.04796 & -0.07799 & -0.08883 & -0.08364 & -0.07119 & -0.05872 & -0.04563 & -0.03189 & -0.01680 \\
\cline{3-13}
 &50&-0.22&   & -0.00465 & -0.01648 & -0.02828 & -0.03997 & -0.05214 & -0.06481 & -0.07481 & -0.07564 & -0.05945 \\
&& -0.11&   & -0.00341 & -0.00992 & -0.01641 & -0.02275 & -0.02925 & -0.03598 & -0.04128 & -0.0416 & -0.03277 \\
&&0& $ G$  & 0.00004 & -0.00360 & -0.00870 & -0.01420 & -0.02003 & -0.02571 & -0.02945 & -0.02849 & -0.02014 \\
&&0.22&  & 0.00423 & 0.01059 & 0.01672 & 0.02258 & 0.02844 & 0.03423 & 0.03724 & 0.03375 & 0.02180 \\
&&0.5&  & 0.01194 & 0.02564 & 0.03830 & 0.05015 & 0.06178 & 0.07333 & 0.07874 & 0.07006 & 0.04463 \\
&&0.9&& 0.01686 & 0.03138 & 0.04476 & 0.05740 & 0.06988 & 0.08231 & 0.08741 & 0.07604 & 0.04647 \\
\cline{2-13}
Frank&&-0.22&  & 0.04584 & 0.06426 & 0.06634 & 0.05840 & 0.04727 & 0.03642 & 0.02544 & 0.01466 & 0.00401 \\
&&-0.11&  & 0.02302 & 0.03241 & 0.03356 & 0.02981 & 0.02432 & 0.01878 & 0.01330 & 0.00787 & 0.00260 \\
&&0&$ F$  & 0.01719 & 0.02537 & 0.02740 & 0.02478 & 0.02018 & 0.01566 & 0.01117 & 0.00669 & 0.00209 \\
&&0.22& & -0.02421 & -0.03686 & -0.04050 & -0.03699 & -0.03057 & -0.02404 & -0.01753 & -0.01083 & -0.00411 \\
&&0.5&  & -0.04747 & -0.07421 & -0.08263 & -0.07646 & -0.06401 & -0.05133 & -0.03819 & -0.02478 & -0.01084 \\
&&0.9&& -0.05041 & -0.07994 & -0.09032 & -0.08406 & -0.07071 & -0.05722 & -0.04335 & -0.02927 & -0.01414 \\
\cline{3-13}
&100&-0.22& & -0.00402 & -0.01476 & -0.02564 & -0.03669 & -0.04786 & -0.05902 & -0.06733 & -0.06604 & -0.04860 \\
&&-0.11& & -0.00257 & -0.00796 & -0.01339 & -0.01886 & -0.02443 & -0.02994 & -0.03385 & -0.03288 & -0.02382 \\
&&0&$ G$ & -0.00024 & -0.00411 & -0.00912 & -0.01458 & -0.02027 & -0.02566 & -0.02880 & -0.02719 & -0.01912 \\
&&0.22& & 0.00410 & 0.01083 & 0.01738 & 0.02384 & 0.03029 & 0.03668 & 0.04020 & 0.03649 & 0.02368 \\
&&0.5&  & 0.01078 & 0.02457 & 0.03767 & 0.05051 & 0.06323 & 0.07576 & 0.08200 & 0.07329 & 0.04674 \\
&&0.9& & 0.00067 & 0.00100 & 0.00130 & 0.00164 & 0.00199 & 0.00249 & 0.00290 & 0.00278 & 0.00196 \\
   \hline
& & 0&    & 0.00164 & 0.00329 & 0.00422 & 0.00464 & 0.00502 & 0.00450 & 0.00402 & 0.00310 & 0.00173 \\
&&0.22&$ F$  & -0.05036 & -0.07621 & -0.08337 & -0.07646 & -0.06415 & -0.05153 & -0.03841 & -0.02500 & -0.01065 \\
&&0.5& & -0.02829 & -0.03898 & -0.04023 & -0.03591 & -0.02958 & -0.02307 & -0.01679 & -0.01010 & -0.00348 \\
&&0.9& & -0.00704 & -0.00950 & -0.00970 & -0.00855 & -0.00692 & -0.00530 & -0.00361 & -0.00199 & -0.00044 \\
\cline{3-13}
 &50&0&    & 0.00167 & 0.00308 & 0.00430 & 0.00465 & 0.00485 & 0.00461 & 0.00413 & 0.00330 & 0.00182 \\
&& 0.22&$ G$  & 0.00941 & 0.02171 & 0.03378 & 0.04579 & 0.05801 & 0.07070 & 0.07626 & 0.06715 & 0.04175 \\
&&0.5&   & 0.00636 & 0.01285 & 0.01915 & 0.02534 & 0.03143 & 0.03766 & 0.03927 & 0.03327 & 0.02001 \\
&&0.9& & 0.00163 & 0.00351 & 0.00520 & 0.00682 & 0.00836 & 0.00988 & 0.01022 & 0.00853 & 0.00510 \\
\cline{2-13}
Clayton&&0&  & 0.00088 & 0.00159 & 0.00202 & 0.00246 & 0.00245 & 0.00231 & 0.00197 & 0.00147 & 0.00087 \\
&&0.22&$ F$    & -0.00651 & -0.00863 & -0.00866 & -0.00754 & -0.00608 & -0.00463 & -0.00322 & -0.00182 & -0.00039 \\
&&0.5&& -0.03055 & -0.04118 & -0.04195 & -0.03709 & -0.03036 & -0.02380 & -0.01704 & -0.01029 & -0.00339 \\
&&0.9&  & -0.05139 & -0.07809 & -0.08441 & -0.07673 & -0.06382 & -0.05058 & -0.03742 & -0.02394 & -0.01005 \\
\cline{3-13}
&100&0& &0.00090 & 0.00156 & 0.00203 & 0.00232 & 0.00243 & 0.00241 & 0.00212 & 0.00163 & 0.00094 \\
&&0.22&$ G$  & 0.00113 & 0.00271 & 0.00421 & 0.00567 & 0.00709 & 0.00846 & 0.00875 & 0.00726 & 0.00428 \\
&&0.5&  & 0.00544 & 0.01221 & 0.01889 & 0.02555 & 0.03213 & 0.03865 & 0.04036 & 0.03398 & 0.02043 \\
&&0.9& & 0.00961 & 0.02258 & 0.03564 & 0.04825 & 0.06103 & 0.07388 & 0.07959 & 0.06966 & 0.04333 \\ \hline
\end{tabular}}
\end{table}

\end{document}